\documentclass[10pt,journal,compsoc]{IEEEtran} 
\ifCLASSOPTIONcompsoc
  \usepackage[nocompress]{cite}
\else
  \usepackage{cite}
\fi

\usepackage[utf8]{inputenc}
\usepackage[T1]{fontenc}

\usepackage{mathptmx} %

\usepackage{amsmath} %
\usepackage{amssymb}  %
\usepackage{bm}
\usepackage{graphicx} %
\usepackage{subcaption}
\usepackage{hyperref}
\usepackage{float}
\usepackage{multicol}
\usepackage{lipsum}
\usepackage{mwe}

\title{
Multi-Modal Super Resolution for Dense Microscopic Particle Size Estimation
}

\author{Sarvesh Patil\IEEEauthorrefmark{1}, Chava Y P D Phani Rajanish\IEEEauthorrefmark{1}, and Naveen Margankunte

\IEEEcompsocitemizethanks{\IEEEcompsocthanksitem Sarvesh Patil and Chava Y P D Phani Rajanish contributed equally to this work.
\IEEEcompsocthanksitem Sarvesh Patil, Chava Y P D Phani Rajanish, Naveen Margankunte are with HyperWorks Imaging, Bangalore, India.

\protect

E-mail: $\left\{ sarvesh,\: rajanish,\: naveen \right\}$ @hyperworksimaging.com }
}

\begin{document}

\IEEEtitleabstractindextext{
\begin{abstract}

Particle Size Analysis (PSA) is an important process carried out in a number of industries, which can significantly influence the properties of the final product. A ubiquitous instrument for this purpose is the Optical Microscope (OM). However, OMs are often prone to drawbacks like low resolution, small focal depth, and edge features being masked due to diffraction. We propose a powerful application of a combination of two Conditional Generative Adversarial Networks (cGANs) that Super Resolve OM images to look like Scanning Electron Microscope (SEM) images. We further demonstrate the use of a custom object detection module that can perform efficient PSA of the super-resolved particles on both, densely and sparsely packed images. The PSA results obtained from the super resolved images have been benchmarked against human annotators, and results obtained from the corresponding SEM images. The proposed models show a generalizable way of multi-modal image translation and super-resolution for accurate particle size estimation.

\end{abstract}

\begin{IEEEkeywords}
Generative adversarial networks, object detection, microscopic particle size analysis, neural networks
\end{IEEEkeywords}
}

\maketitle
\thispagestyle{empty}
\pagestyle{empty}
\section{INTRODUCTION}

Particle Size Analysis (PSA) enables accurate estimation of particle sizes of microspheres, offering insights into the material or the powder being formed, and is often a crucial process in manufacturing pipelines. Estimating the size of particles in a sample can help in determining many physical and chemical properties of the material like packing density, flow characteristics, reactivity, dissolution rate, texture, and appearance of the finished product \cite{Fu2008}\cite{Laceby2017}.

The most common techniques for PSA are Laser Diffraction (LD), Dynamic Light scattering (DLS) and Imaging Particle Analysis (IPA). IPA is extensively used For microspheres where particle shape and size is of extreme importance. A significant advantage of using OM image-based analysis is that, it is a non-intrusive method with reasonable costs and decent performance. However, more advanced microscopy techniques like Scanning Electron Microscopes (SEMs) that are quite expensive can be used for granular magnification to capture images with sharp boundaries and depth information.

\begin{figure}[t]
    \centering
      \begin{subfigure}[b]{0.45\linewidth}
        \includegraphics[width=\linewidth]{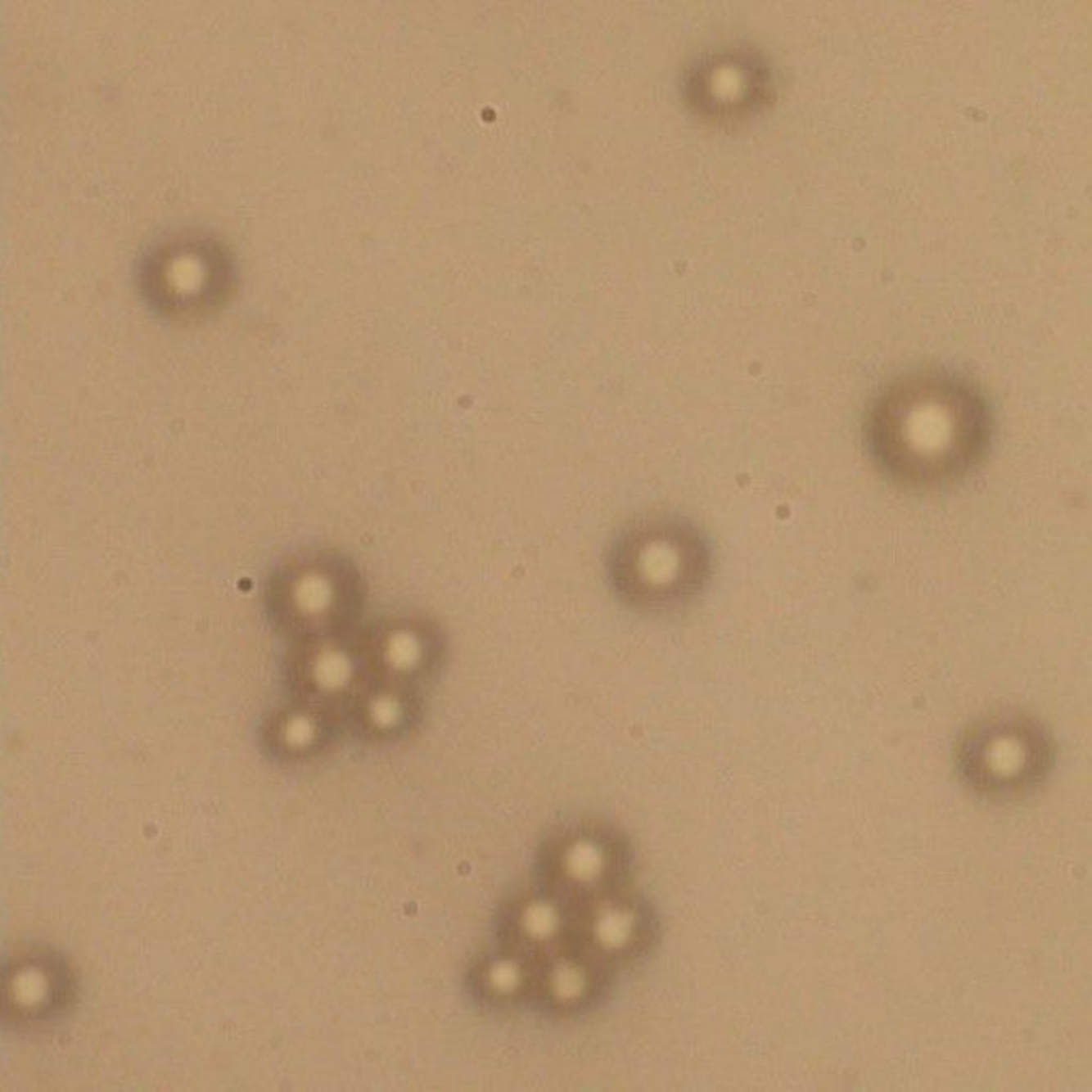}
        \caption{OM}
        \label{fig:blur_OM}
      \end{subfigure}
      \begin{subfigure}[b]{0.45\linewidth}
        \includegraphics[width=\linewidth]{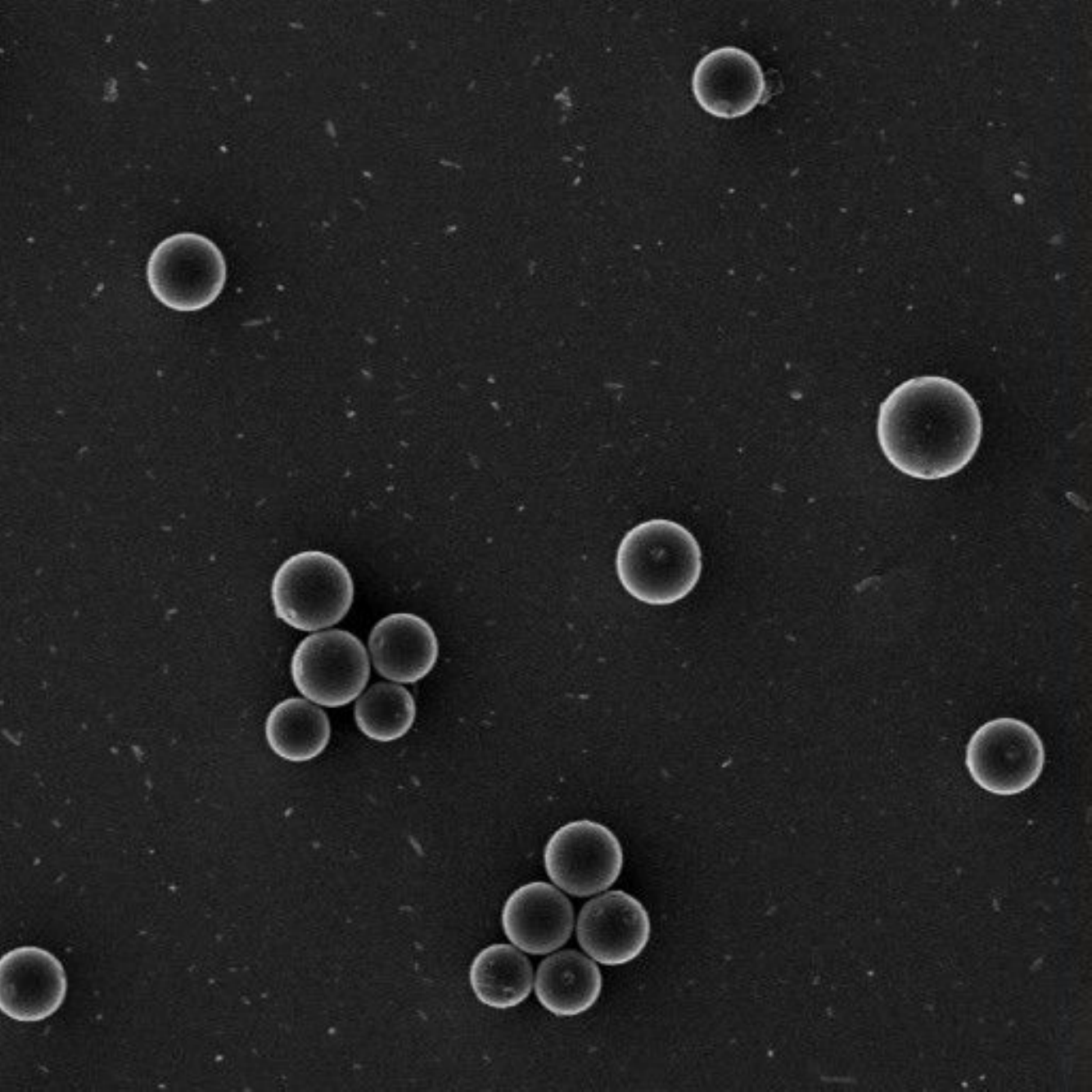}
        \caption{SEM}
        \label{fig:sharp_SEM}
      \end{subfigure}
    \caption{Ground truth images. (a) shows bleeding of edges in OM images. (b) shows sharp edges from SEM images. }
    \label{fig:omsem}
\end{figure}

The use of traditional Image processing techniques like edge extraction, thresholding, and segmentation, for PSA, has been shown in \cite{Wang2006}\cite{Vega2015}\cite{Chen2013}. The segmentation maps obtained are used for particle sizing. A conversion factor is used to obtain real values using the size of the particles in pixels.

Methods like Hough Circle Transform (HCT) have been used for localization and counting of cell nuclei in phase-contrast microscopy images \cite{Smereka2008}. In \cite{Mirzaei2017}, a modified HCT was used for nanoparticle diameter estimation. Although HCT performs well for high-resolution electron microscopy images, results on OM images need parameter fine-tuning due to variable lighting and background conditions across various images. This reduces the accuracy of predictions of the algorithm for unseen images. 

Deep neural network based methods have proven to be robust, and outperform the traditional image processing based techniques \cite{Zhao2019}. We use a Convolutional Neural Network (CNN) in the form of a custom object detection model to perform particle size analysis. Use of CNNs for cell nuclei classification, detection, and segmentation of nanoparticles has been shown in \cite{Zhu2017}\cite{Wang2016}\cite{Xiao2017}\cite{Gueven2018}. These approaches have proven useful for particle sizing on high-quality SEM images. However, when applied on OM images, these methods fail to detect dense agglomerates of blurred particles effectively. 	

Although use of deep neural network based approaches seems straightforward, performance of these models is heavily dependent on the ground truths being perfect. However, accurate annotation of optical microscopy particles is a non-trivial task, even for human annotators. Due to bleeding of the edges of the particles caused by diffraction of light, accurately sizing them is an uphill task. On the other hand, from Fig. \ref{fig:omsem}, we can see that annotations for SEM images can be easily created due to their sharp edges. Thus we propose that the quality of the ground truth data can be improved significantly by converting OM images to the SEM domain and then applying an object detection algorithm to perform particle size analysis. 

To convert OM images to look like SEM images, image-based conditional Generative Adversarial Networks (GANs) like Pix2Pix \cite{Zhu2018} can be used. This process is known as image translation, and it converts images from OM domain to the SEM domain. However, the Pix2Pix model fails to add sharp edges and detailed texture information of the SEM particles. Hence, we use the ESRGAN model \cite{Lim2017} for the super-resolution of the Pix2Pix output. In this regard, the core contributions of this work are as follows:

\begin{itemize}
    \item A combination of two GANs that can be used to translate and subsequently Super Resolve OM images to look like SEM images.
    \item A dense object detection model to perform PSA on super-resolved images.
\end{itemize}

\section{BACKGROUND}

\subsection{Super Resolution GANs} 
There has been great improvement in the quality of images generated by GANs since their inception. However, for the most part, GANs are considered as black boxes with a utopian Nash Equilibrium that may never be reached \cite{Farnia2020}. Despite being difficult to train, the synthesizing capability of GANs has proven useful in myriads of applications \cite{Karras2018}\cite{Wang2020}. Conditional GANs (cGANs) \cite{mirza2014conditional} demonstrate the capability to condition the Generator to produce desired output. cGANs have made great progress in image generation from text \cite{Reed2016}, image in-painting \cite{Pathak2016}, style transfer \cite{Zhu2018} and super-resolution \cite{Ledig2017}. 

There have been many works presenting the use of GANs \cite{Goodfellow2014} to super resolve microscopy images \cite{Haan2019}\cite{Wang2019}\cite{GrantJacob2019}. In \cite{Wang2019}, cGANs have been used for cross-modal super-resolution of Stimulated Emission Depletion (STED) microscopy images. In \cite{Haan2019}, a U-Net based GAN was used for the super-resolution of low-resolution SEM images. These methods involve the use of images from similar distribution to perform super-resolution. 

In \cite{GrantJacob2019} segmentation maps of OM images of pollen grains were super resolved to look like SEM images at a higher magnification. This approach shows use of multi-modal super resolution. However, the amount of individual preprocessing involved in segmenting each particle is not practical for our application, due to the presence of a large number of clustered particles. Hence, we propose use of a Pix2Pix model for the initial image translation and an ESRGAN model for the subsequent super resolution. Our method is reproducible and works without the need of manual preprocessing on the particles.

\subsection{Particle Size Analysis} 
Some novel techniques in PSA involve the use of neural networks for regression and object detection. In \cite{Hamzeloo2014}, neural networks have been used to estimate sieve size for particles on a conveyor belt, based on edge detection metrics like Feret distance, best-fit rectangle, area, etc. In \cite{Frei2020}, the use of Mask R-CNN has been demonstrated for object detection, segmentation, and particle size analysis. These approaches have proven useful for high-quality electron microscopy images. However, their use on optical microscope images becomes a difficult task, primarily due to the thick blurred edges.

From the region-based object detection networks to monolithic object detection networks like YOLO \cite{Redmon2016}, RetinaNet \cite{Lin2017} and SSD \cite{Liu2016}, the field has witnessed an astonishing growth in algorithms in the last few years. Single shot detection algorithms have dominated the field with high mean Average Precision (mAP) values while being able to run in real-time. However, even with dense object detection algorithms like RetinaNet, these algorithms often find it difficult to detect and size densely packed particles with high precision. 

Our object detection algorithm also uses a single feed-forward operation but has a four-dimensional patch at its output as opposed to a fully connected layer. Also, since our microscopic particles are circular in nature, we regress their circle coordinates $(x, y, r)$ and confidence $p$ as opposed to box coordinates. This method has been inspired by \cite{Sam2020}, where the authors present a multi-scale approach for object detection in high-density images. 

\begin{figure}[ht]	
    \centering	
      \begin{subfigure}[b]{0.8\linewidth}	
        \includegraphics[width=\linewidth]{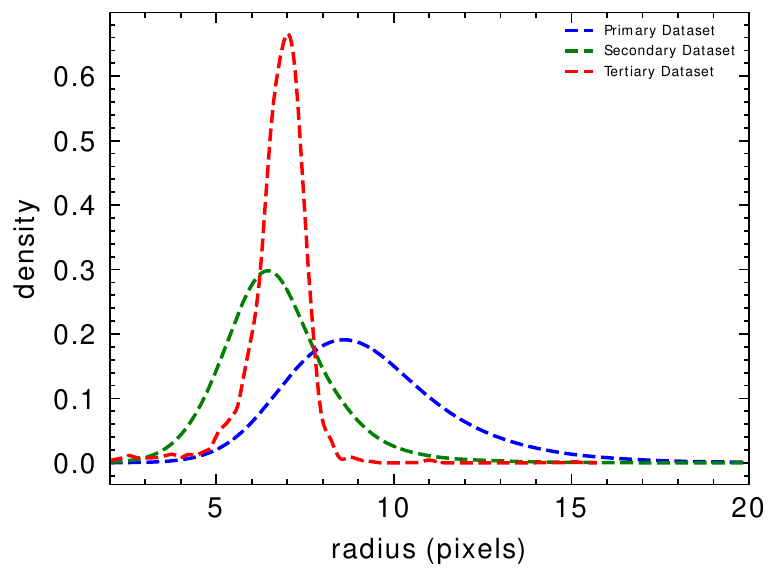}	
      \end{subfigure}	
    \caption{Primary, Secondary and Tertiary data set distributions}
    \label{fig:data_distribution}	
\end{figure}

We devise an object detection algorithm -- Super Resolution Particle Sizing Algorithm (SR-PSA) -- which is used to detect particles in sparse as well as highly dense clusters of particles. Along with our "primary" dataset, which is used for preliminary training and validation, we use two more datasets -- "secondary" and "tertiary", to measure the performance of the model on particles with varied distributions as shown in Fig. \ref{fig:data_distribution}. We compare the results from our algorithm to those generated by human annotators.

\section{METHOD}

\begin{figure}[H]
    \centering
      \begin{subfigure}[b]{0.45\columnwidth}
        \includegraphics[width=\linewidth]{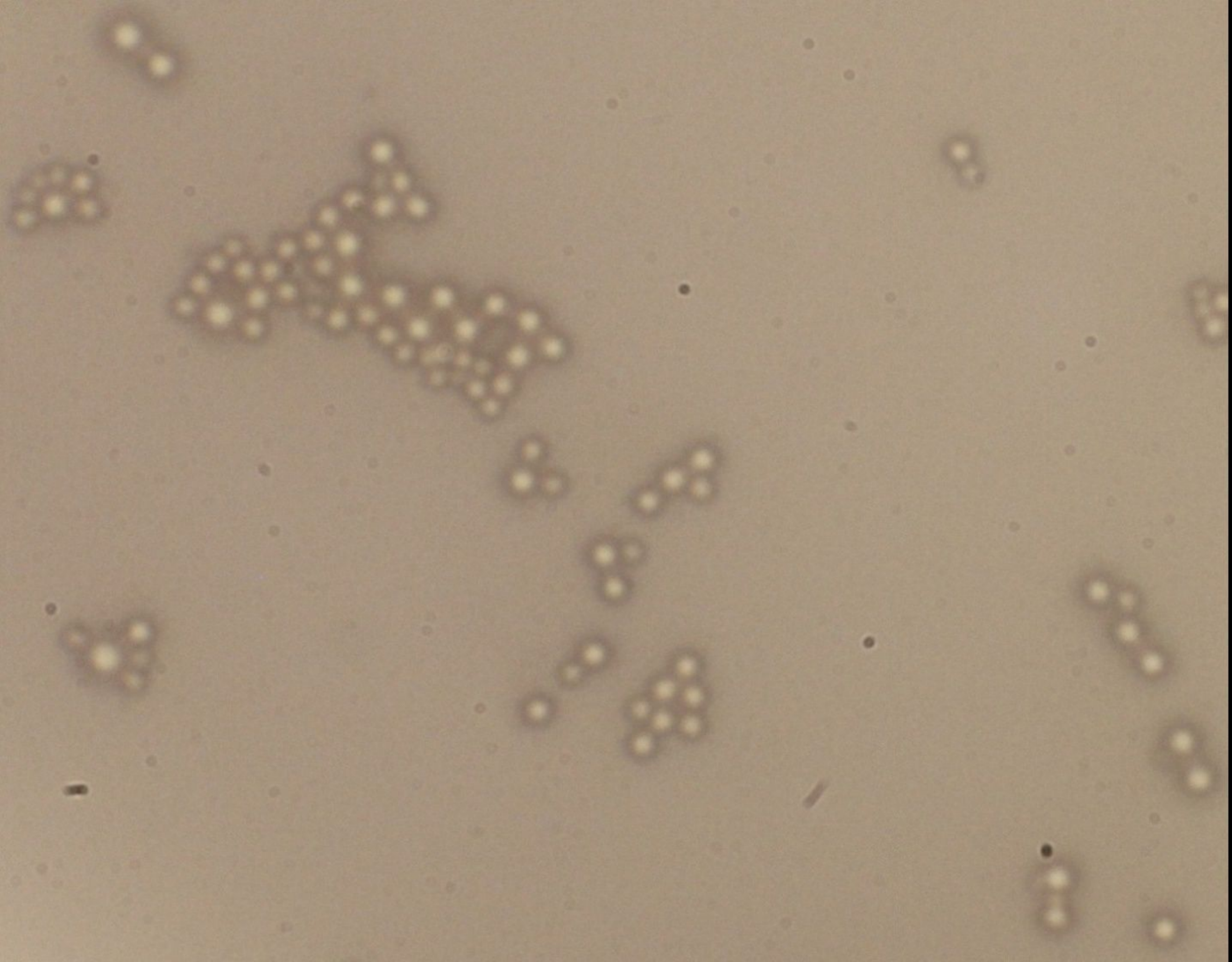}
        \caption{OM}
        \label{fig:OM_a}
      \end{subfigure}
      \begin{subfigure}[b]{0.45\columnwidth}
        \includegraphics[width=\linewidth]{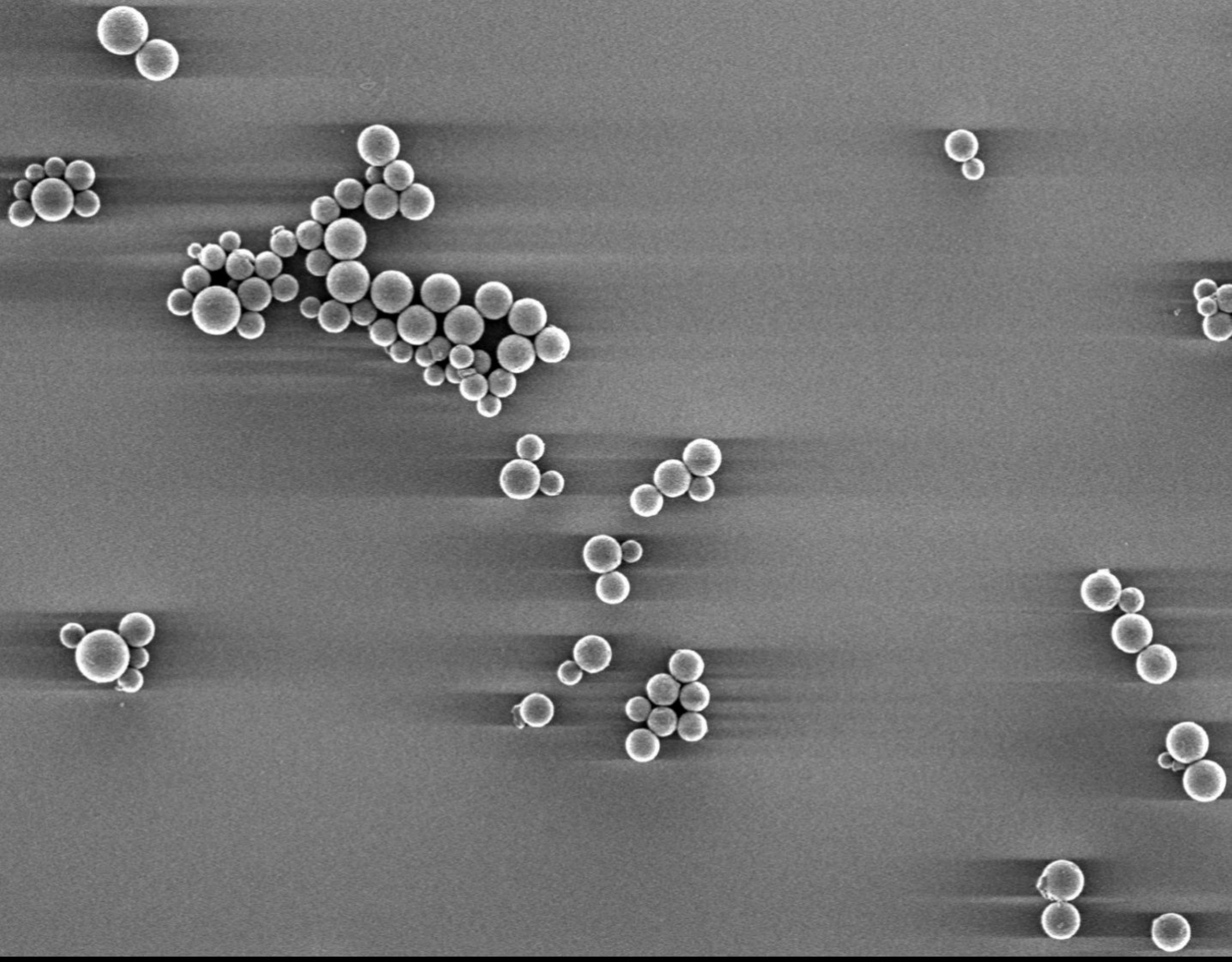}
        \caption{SEM}
        \label{fig:SEM_b}
      \end{subfigure}
      \begin{subfigure}[b]{0.45\columnwidth}
        \includegraphics[width=\linewidth]{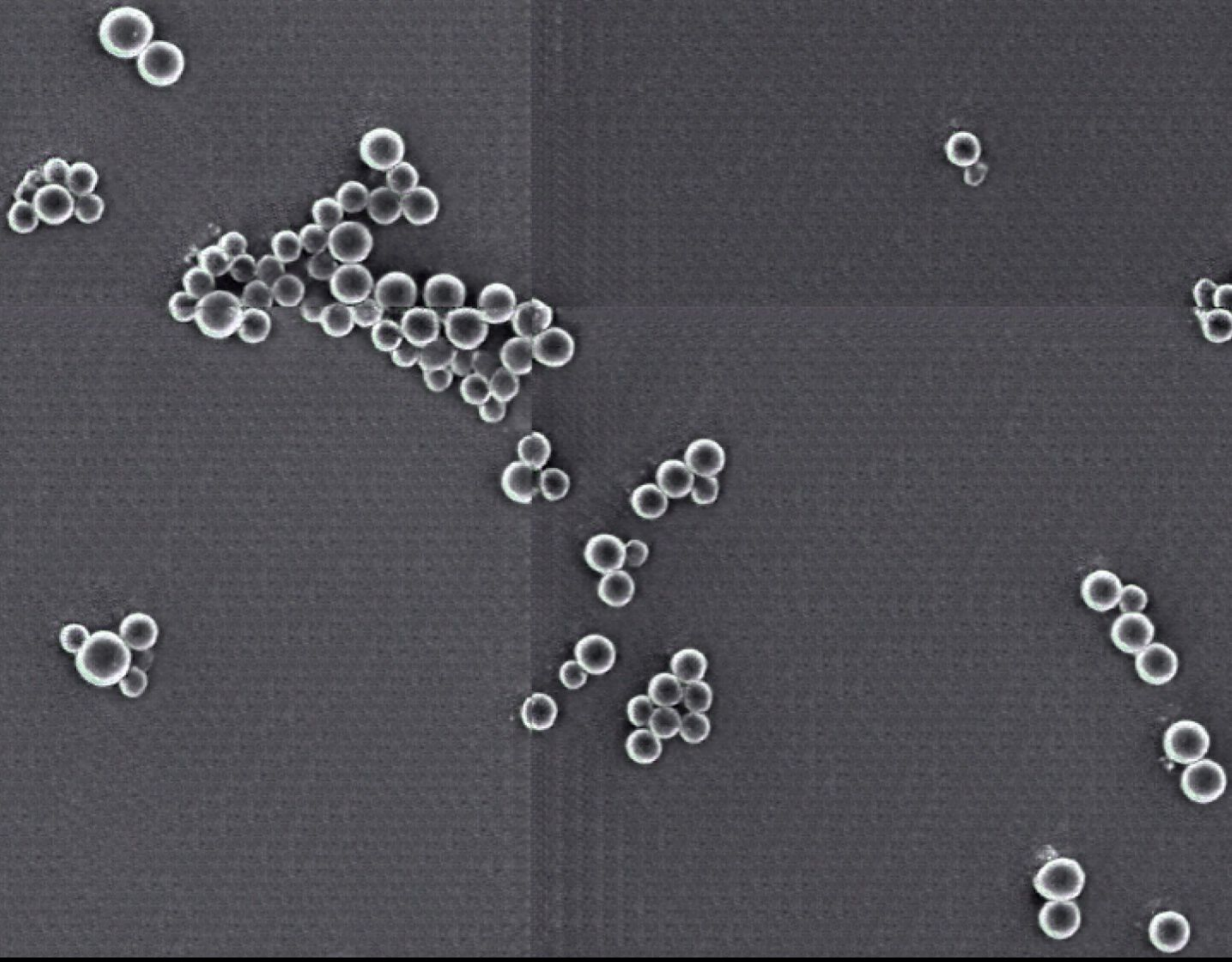}
        \caption{SR}
        \label{fig:SR_c}
      \end{subfigure}
      \begin{subfigure}[b]{0.45\columnwidth}
        \includegraphics[width=\linewidth]{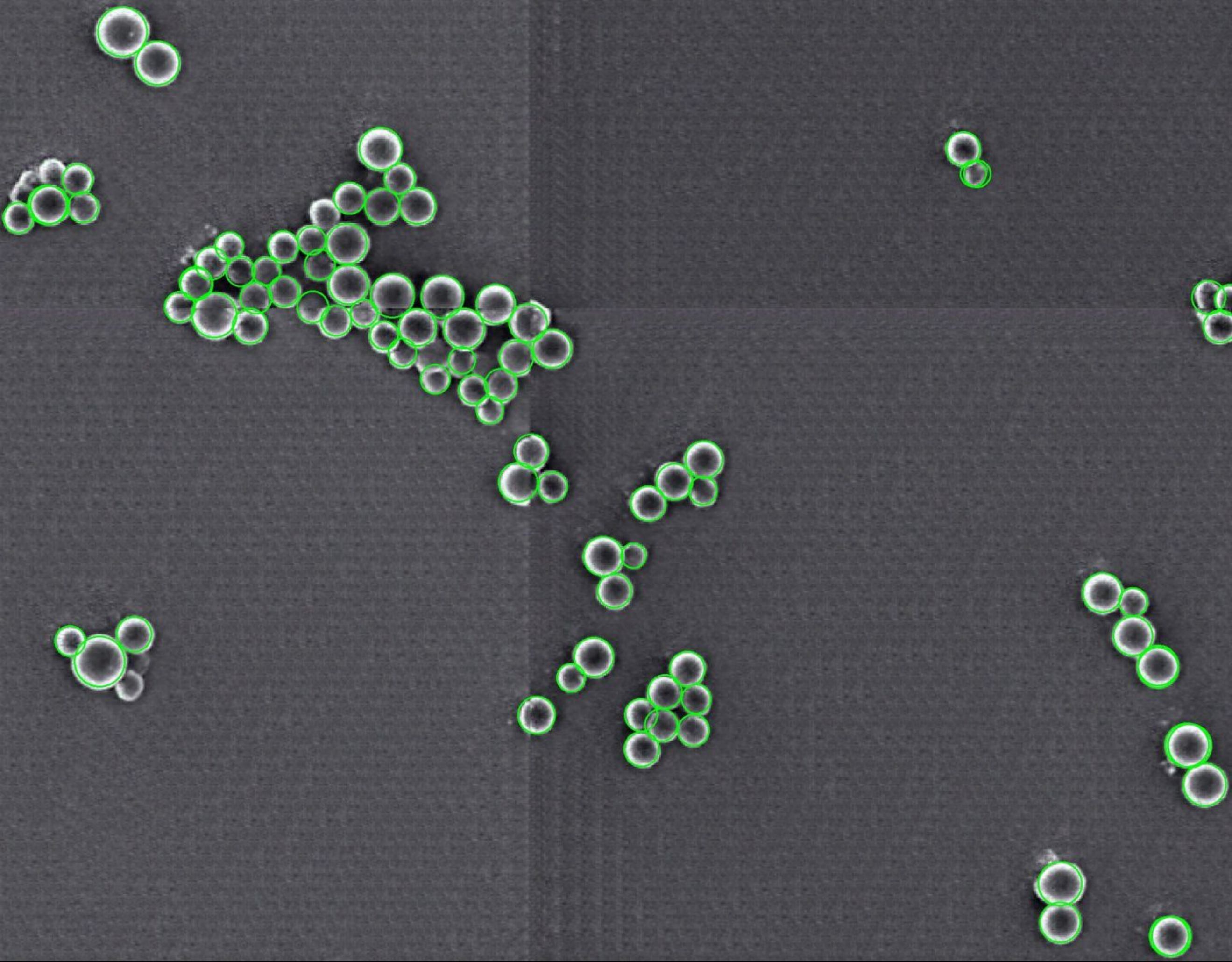}
        \caption{SR with PSA}
        \label{fig:SR_d}
      \end{subfigure}
    \caption{(a), (b) are the ground truth images, (c) is the super resolved image, and (d) shows PSA output on (c).}
    \label{fig:fourimgs}
\end{figure}

\begin{figure*}
  \includegraphics[width=\textwidth]{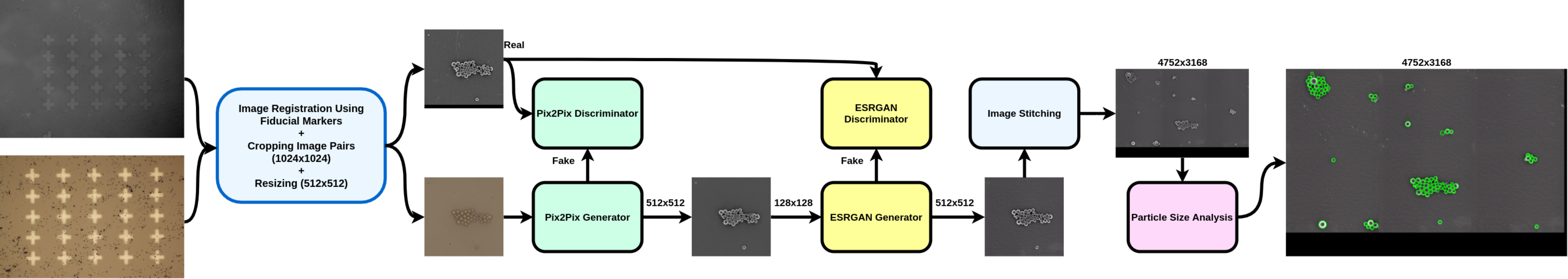}
  \caption{The pipeline describing the flow of the entire process}
  \label{fig:pipeline}
\end{figure*}
\subsection{Image Pair Registration}
Image Pair Registration is an important step for Super-Resolution using GANs, as it aligns the image in one mode (OM) with respect to another (SEM). This enables the model to learn the one-to-one mapping between the low-resolution OM images and the high-resolution SEM images. As shown in Fig. \ref{fig:pipeline} OM and SEM image pairs were obtained by using fiducial markers on the captured samples. The image pairs were registered using affine transforms in a landmark-based registration algorithm. The semi-automatic algorithm used landmark points from the user on the input image and the target image to perform registration.

\subsection{Multimodal Image Translation}
We use the Pix2Pix GAN \cite{Isola2018} for multimodal image translation as our baseline model. In the Pix2Pix architecture, a conditional image is provided as input to the U-Net generator. The down-sampling layers in the U-Net model form a latent representation of the input image. By using advanced operations like Spectral Normalization \cite{Miyato2018}, Self-Attention \cite{Zhang2019}, and Anti-Aliasing pooling \cite{Zhang2019a} on the convolutional layers, we improve the latent representations of our input data. This improves the performance of the generator by simplifying the task for the up-sampling layers, thus enabling faster convergence of the model. We also implement spectral normalization and anti-aliasing pooling for the PatchGAN discriminator \cite{Li2016}.

\textbf{Loss Function} For image-conditional GANs, the generator loss is obtained by calculating the error between the generated output and the real image. L1 loss function minimizes the sum of the absolute difference between the generated and the target images, while the L2 loss function minimizes the sum of squared differences between the generated and the target images.

L2 loss has been shown to induce blurriness in convolutional generative models due to its averaging effect \cite{Lotter2017}\cite{Pathak2016}. Hence, the default Pix2Pix model implements an L1 loss for the generator. L1 loss promotes sharp boundaries in natural scenes. However, we observe that use of L1 loss results in elongated boundaries for circular particles. Since our main objective is particle sizing for circular particles, these distortions were not acceptable. We propose the use of Root Mean Squared Error (RMSE) loss to mitigate these distortions. RMSE loss increased the circularity of the particles, whilst keeping blurring at a minimum. A simple comparison between the results obtained by these loss functions is demonstrated in Fig. \ref{fig:l1vsl2vsrmse}. Hence, for a $n$ input images $x$ and the generator $G(x)$, our loss is given by:

\begin{equation}
    L_{RMSE}(G) = \mathbb{E}_{x,y}\bigg[\sqrt{\frac{\sum_{i=1}^n (y-G(x))^2}{n}} \bigg]
\end{equation}

Where $y$ is the target SEM image. During the initial epochs, the generator output results in large loss values, which is limited by the RMSE loss function. As training progresses, the losses decrease to small values, which are slightly amplified by the RMSE loss function. This results in a more dynamic weight update as compared to L1 and L2 losses. 

The final optimization over the network is given by:
\begin{equation}
    \arg\min_G \max_D L_{cGAN}(G,D) + \lambda_{RMSE} L_{RMSE}(G)
\end{equation}
Where $\lambda$ controls the weight of RMSE loss function and can be considered as a hyperparameter.

\subsection{Image Super Resolution}
Since Pix2Pix is an image translation model, the output lacks the intricate details of the SEM particles. To add sharp edges and diverse texture information we use another cGAN for super-resolution of the Pix2Pix output. The ESRGAN \cite{Lim2017} model has proven useful in many applications like the super-resolution of old games, low-resolution photographs, old movies, etc. Hence, we train an ESRGAN for our application due to its generalizability across various domains.

The use of Relativistic average discriminator \cite{JolicoeurMartineau2018} ($D_{Ra}$) has been shown to improve the generator by making the discriminator predict relativistic probabilities. Rather than classifying images as real or fake, $D_{Ra}$ tries to predict the probability that a real image $x_r$ is more realistic than a fake image $x_f$ as shown in \cite{Lim2017}.

\textbf{Loss Function} For the generator firstly we use L1 loss between the fake and the real images, for maximizing the sharpness of the particles. The relativistic discriminator loss $L_{GAN}^R$ is taken as the average value of loss generated by the real image, and the loss generated by the fake image \cite{Lim2017}. We also incorporate perceptual loss from the default implementation. A pretrained VGG-19 \cite{Simonyan2015} model is used to extract pre-activation feature maps from the generated output as well as the real image. The L1 distance between the two feature maps is used to calculate the perceptual loss $L_{perceptual}$. Thus, the total generator loss $L_G$ is given as,
\begin{equation}
    L_G = L_{perceptual} + \lambda L_{GAN}^R + \eta L_1
\end{equation}
$\lambda$ and $\eta$ are parameters that can be tuned to give weightage to relativistic loss and L1 loss respectively. As the perceptual loss values are very small, we use $\lambda=0.05$ and $\eta=0.01$. We demonstrate good quality of images being generated by the ESRGAN after certain improvements in Fig. \ref{fig:SR_c}. Generated images preserve the geometrical and structural properties of particles of various sizes through the super-resolution process.

\begin{figure*}[ht!]
      \begin{subfigure}[b]{0.19\textwidth}
        \includegraphics[width=\linewidth]{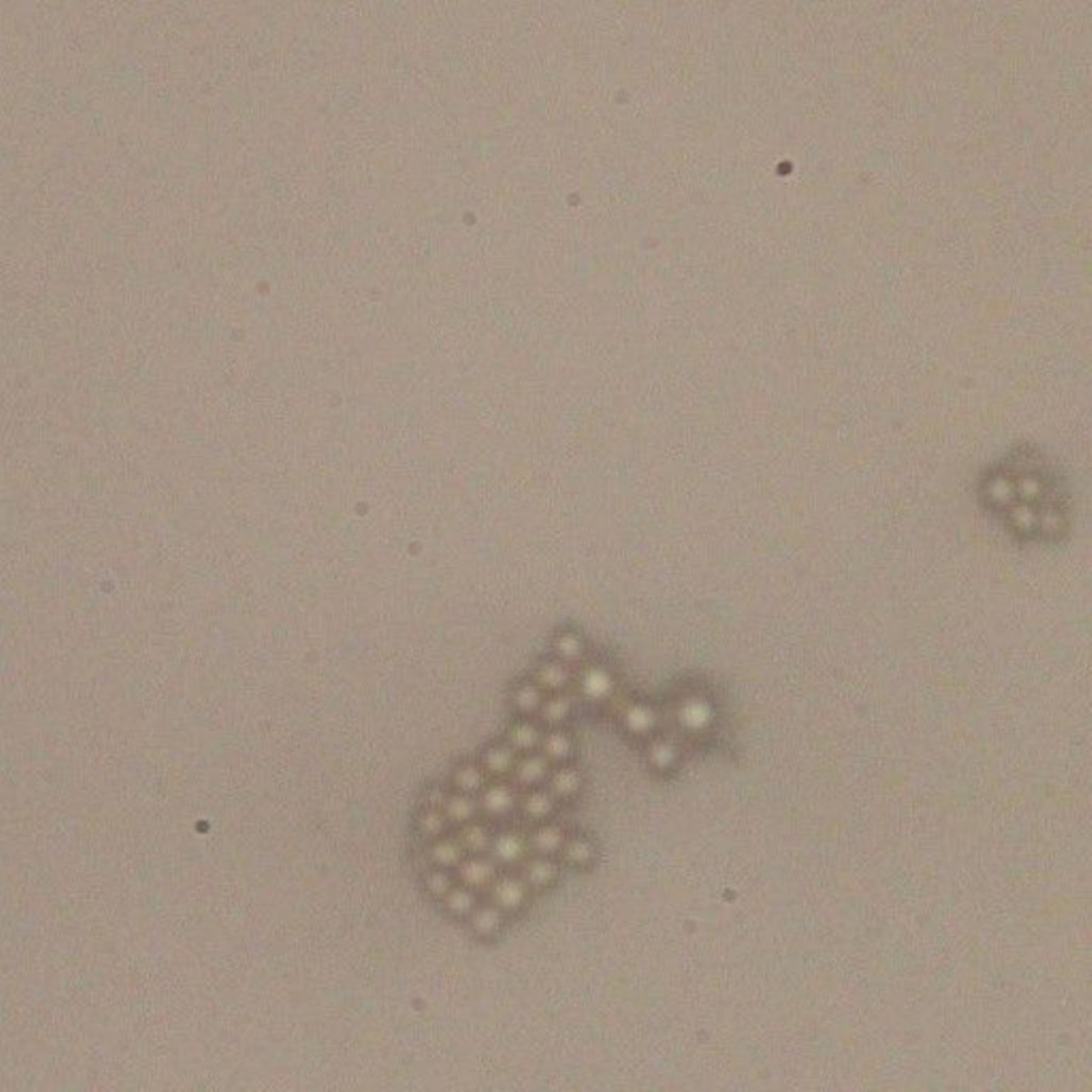}
        \caption{OM}
        \label{fig:compa}
      \end{subfigure}
      \begin{subfigure}[b]{0.19\textwidth}
        \includegraphics[width=\linewidth]{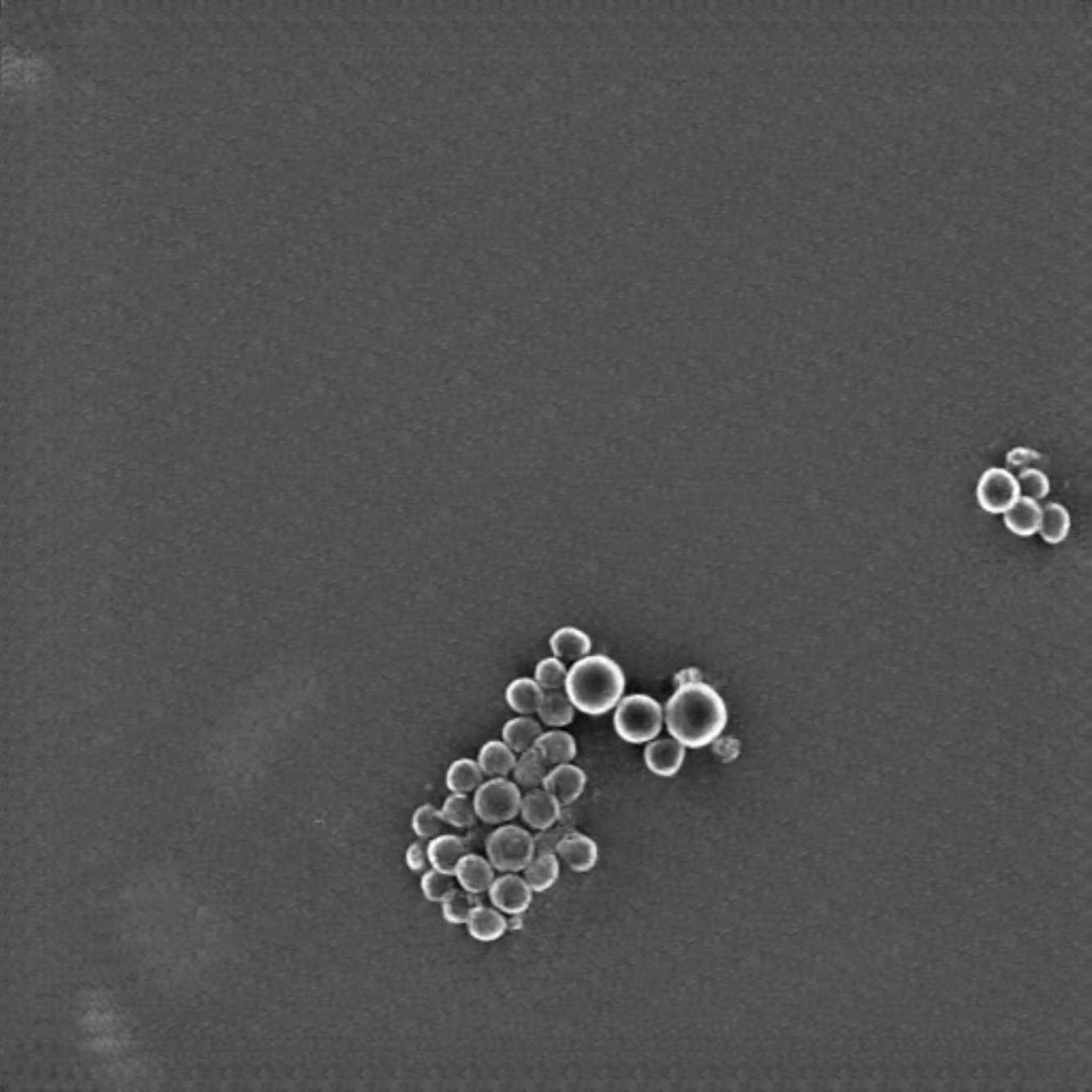}
        \caption{SR}
        \label{fig:compb}
      \end{subfigure}
      \begin{subfigure}[b]{0.19\textwidth}
        \includegraphics[width=\linewidth]{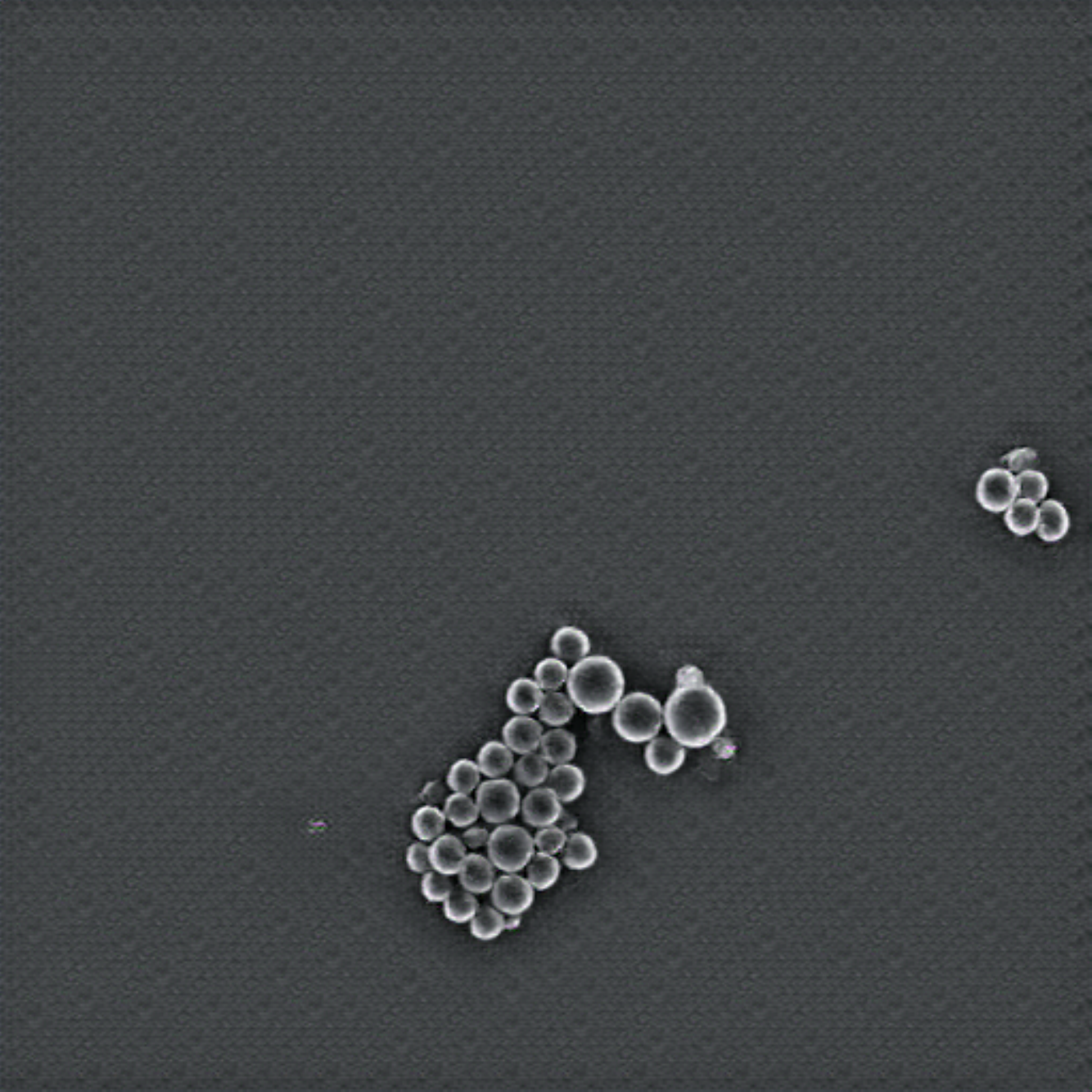}
        \caption{SR + SN}
        \label{fig:compc}
      \end{subfigure}
      \begin{subfigure}[b]{0.19\textwidth}
        \includegraphics[width=\linewidth]{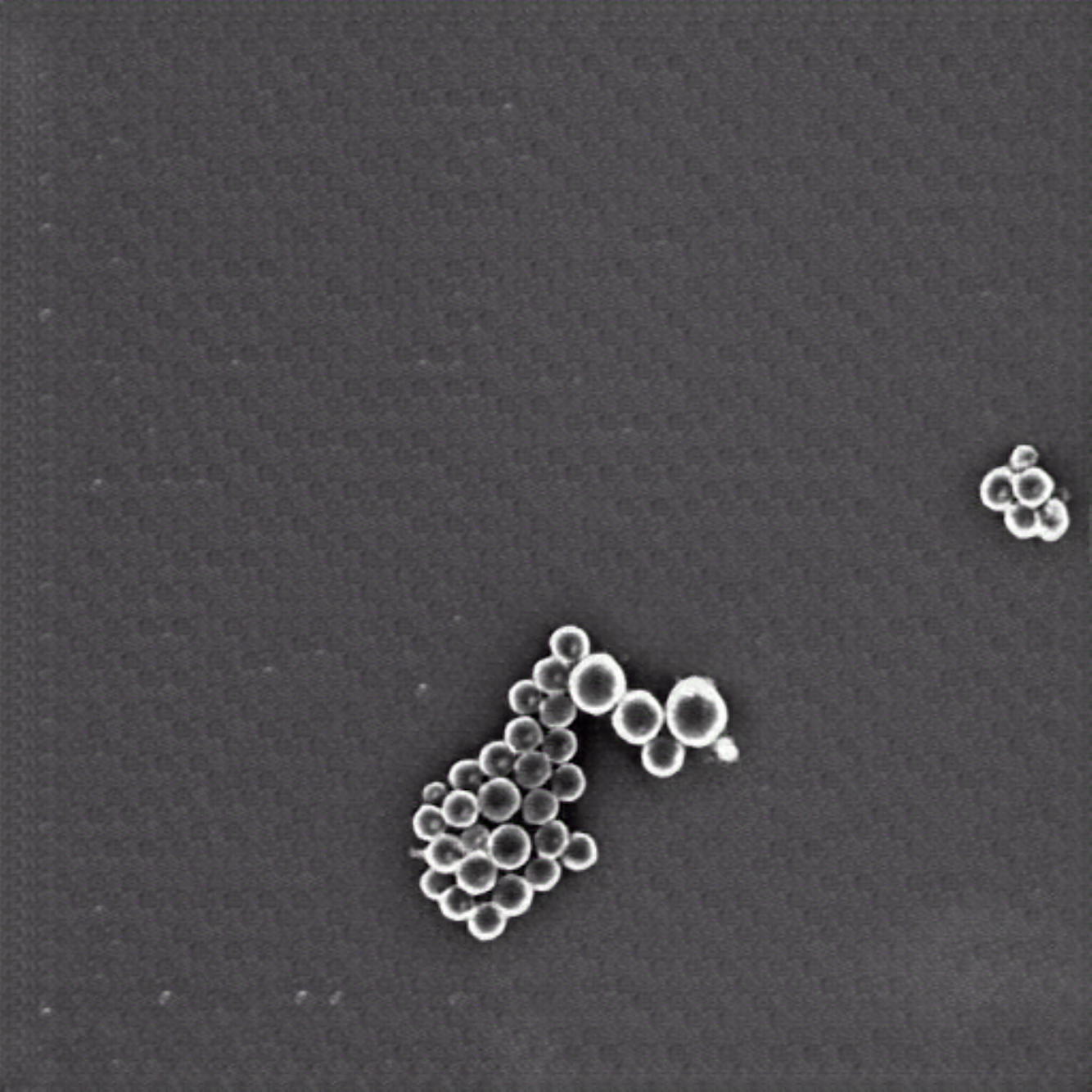}
        \caption{SR + SN + AA + SA}
        \label{fig:compd}
      \end{subfigure}
      \begin{subfigure}[b]{0.19\textwidth}
        \includegraphics[width=\linewidth]{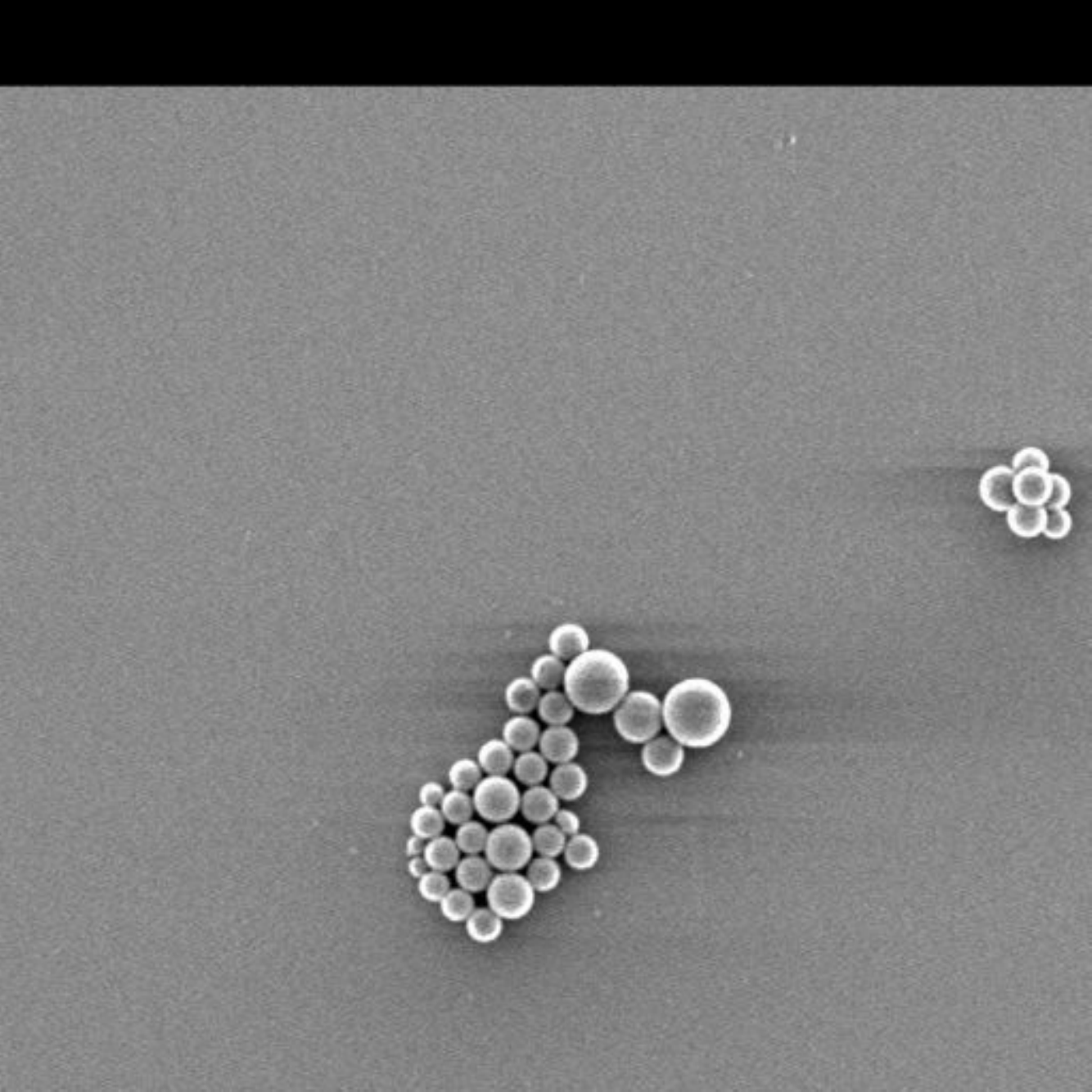}
        \caption{SEM}
        \label{fig:compe}
      \end{subfigure}
    \caption{(a) is the input OM image and (e) is the corresponding SEM image. (b) is the result of default implementation of the pipeline. (c) is the result after adding Spectral Normalization. (d) is the best result obtained after adding all the proposed improvements.}
    \label{fig:improvements}
\end{figure*}

\subsection{Particle Size Analysis} 
For the particle sizing application, instead of bounding boxes, we regress circle coordinates (\emph{x, y}, and \emph{r}). For single class object detection, the need for a separate classification layer is mitigated. Hence, for our application, we use a single feed-forward network to detect circles in the super-resolved images. The ground truth is created using manual annotation over the particles, and the network is trained to predict annotations with high precision and accuracy.

The SR-PSA detection module is a simple Convolutional Neural Network (CNN) that has 6 convolutional layers. The first five layers have a kernel size of $3\times3$ to extract features and the final layer has four feature maps where classification and regression are done in a single iteration. The layers are chosen such that a maximum of two max-pooling operations are performed, i,e the input image is downscaled by 4. The last layer is replaced by a 2D convolution which converts the input feature maps to a 4-dimensional output -- $[p, \:x, \:y, \:r]$, where $p$ is the confidence map and $x$, $y$, and $r$ are the particle coordinates and radius respectively. Since the size of input images is very large ($4752\times3168$), we save the images as $4\times$ and $8\times$ down-scaled images ($1188\times792$, $594\times396$). 

\textbf{Loss Functions} At each pixel level in the feature map, the model is trained to assign high probabilities if the center of a particle exists at that location and low probabilities for any other point in $p$. The probability mask is applied on output to suppress predictions lesser than a predefined threshold ($p_{thresh}$), given as $p_{mask} = p>p_{thresh}$. Similarly, for $x$, $y$, and $r$, the model regresses their respective values at every location in the output. We use L2 distance to compute loss for the model and mask only those regions where the radius values exist. For the $i^{th}$ output, the predicted coordinates are $l$ and ground-truth coordinates are $g$, then the objective is given as,

\begin{equation}
\begin{aligned}
        x_{i_{err}} = \mathbb{E}_{x_i}[MSE(l_x, g_x)\odot p_{mask}] \\
        y_{i_{err}} = \mathbb{E}_{y_i}[MSE(l_y, g_y)\odot p_{mask}] \\
        r_{i_{err}} = \mathbb{E}_{r_i}[MSE(l_r, g_r)\odot p_{mask}] \\
        p_{i_{err}} = \mathbb{E}_{p_i}[\lambda MSE(l_p, g_p)\odot p_{mask}], \\
\end{aligned}
\end{equation}
where $\lambda$ is a tunable weight parameter. The final loss is given as,
\begin{equation}
\label{SRPSA_loss}
    L_{total} = \frac{x_{i_{err}}+y_{i_{err}}+ r_{i_{err}}+p_{i_{err}}}{4}
\end{equation}

\begin{figure*}
  \includegraphics[width=\textwidth]{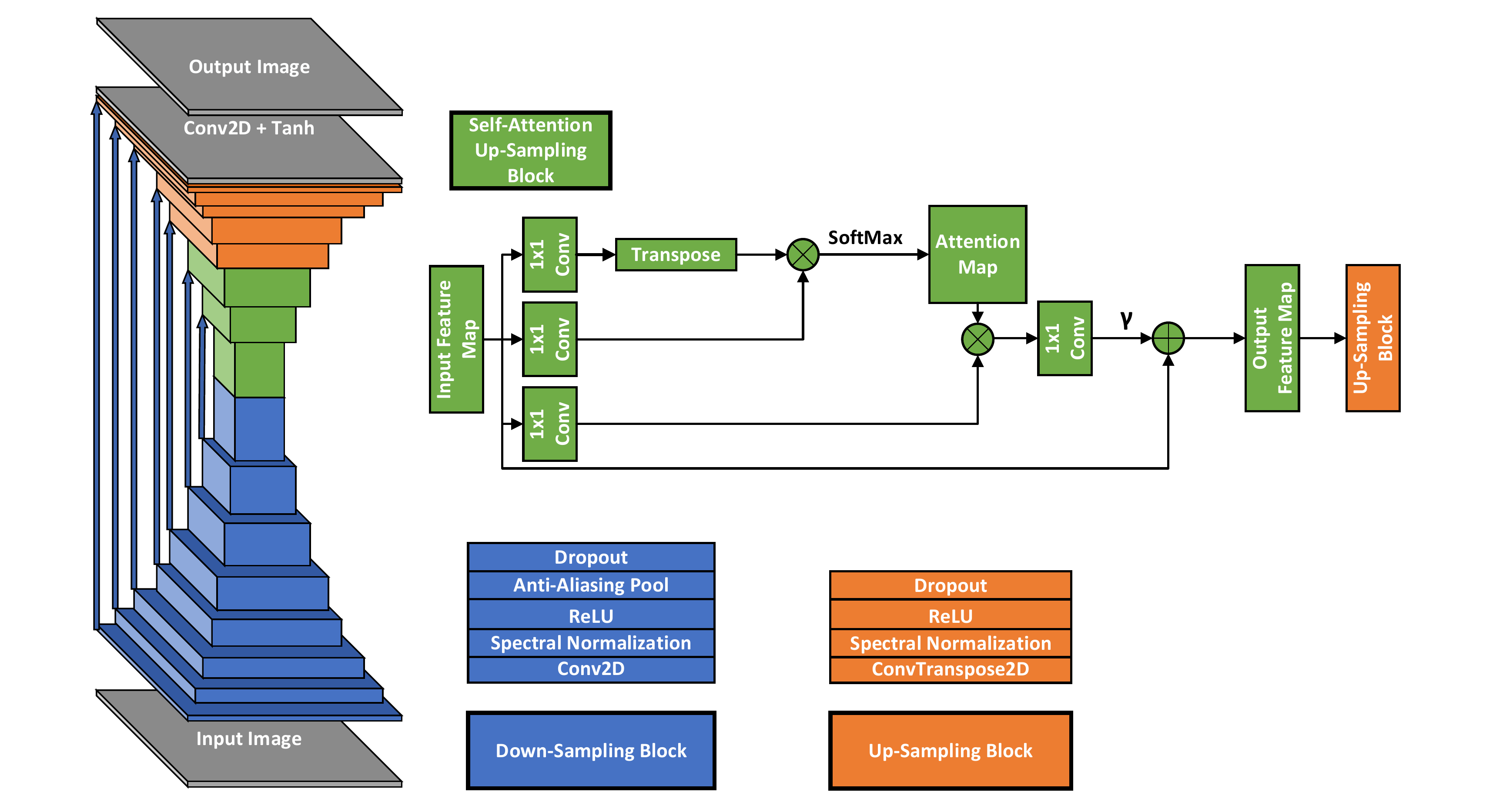}
  \caption{Final Pix2Pix generator architecture after improvements. The self-attention block has been inspired from \cite{Zhang2019a}}
  \label{fig:gan_architecture}
\end{figure*}

\begin{figure*}
  \includegraphics[width=\textwidth]{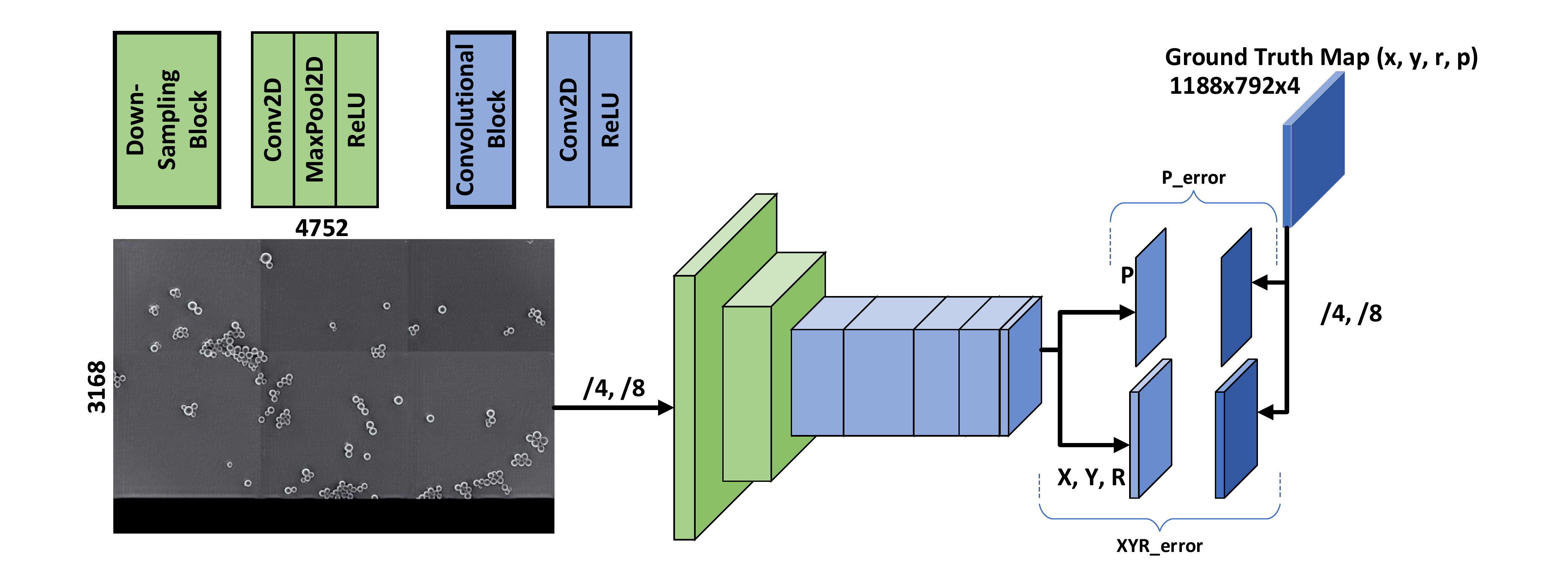}
  \caption{SR-PSA Object Detection model. The CNN has two max-pooling layers. Thus, the ground truth maps are scaled down by 4. For each iteration, the image is down-sampled by 4 and 8 for the same model.}
  \label{fig:psa_architecture}
\end{figure*}

\section{GAN IMPROVEMENTS}
The default implementation (Pix2Pix with the ESRGAN) generated particles with boundaries blended into other particles in the cluster. As a result, the particle sizing algorithm found it difficult to regress circle coordinates, giving incorrect size estimations. Hence, we added the state-of-the-art improvements in CNNs and GANs to our models for better-looking images with sharp and clear boundaries. The changes can be tracked in Fig. \ref{fig:improvements}. 

\subsection{Spectral Normalization} We use Spectral Normalization (SN) \cite{Miyato2018} on all the down-sampling 2D convolution layers of our Pix2Pix as well as ESRGAN models. This helps us bound the latent space generation of the deep feature maps. In the original implementation of ESRGAN, batch normalization was eliminated due to less generalizability and the presence of unpleasant artifacts. But without any form of normalization, our experiments often resulted in either mode-collapse or overfitting. The ESRGAN training stability improved by the use of SN over the generator as well as the discriminator. 

Even with large variations in the input images, the variation in the distribution of generated images remains low due to the addition of SN. In all the experiments done using our model, observed convergence of the losses after incorporating SN. Spectral normalization only requires one hyperparameter tuning, the Lipschitz constant, which works well at its default value of 1. 

A function $f$ is said to be Lipschitz continuous if, for a given metric ($L2$ Loss), it satisfies the following condition:
\begin{equation}
||f(y) - f(x)|| \leq K||y-x||
\end{equation}
If K is minimum, then it is the Lipschitz constant of a function, and Lipschitz continuity limits the rate of change of the function with respect to K. The weights in the convolutional layers can be bound by a Lipschitz constant which is controlled by applying spectral norm on each layer. 

In \cite{Miyato2018} is shown that for a weight matrix $W$, the spectral norm is given by the square root of the largest eigenvalue of $W^TW$. The \emph{Power Iteration} method is used to compute the spectral norm of $W$, given by $\sigma(W)$, efficiently. Spectral Normalization of a layer then becomes simply replacing $W$ by $W/\sigma(W)$.

\subsection{Anti-Aliasing} The down-sampling network in U-Net uses strided-convolutions in the default Pix2Pix implementation. With strided-convolutions, we observe deformations of particle boundaries for images with size distributions different than training images. Hence, we adopt anti-aliasing pooling \cite{Zhang2019} in our Pix2Pix model for the down-sampling layers in the generator as well as the discriminator. We observe increased circularity of the particles with the addition of anti-aliasing pooling, especially in dense particle clusters. Although the anti-aliasing operations require more compute than simple max-pooling operations, the performance increase makes up for it.

\subsection{Self-Attention} With the final addition of Self-Attention (SA), the output from our Pix2Pix model shows uniform illumination over the entire generated image, whilst maintaining high recall of boundaries and edges of the particles. The modified generator architecture is shown in Fig. \ref{fig:gan_architecture}. The number of parameters in SA layers increase exponentially with increase in the size of the feature map. For our application, since the size of input images is $512\times512$, we restrict SA operations over the smallest three layers in the generator.

For a hidden layer, the self-attention \cite{Wang2018} algorithm computes three $1\times1$ convolutions, ${f(x)},\:{g(x)}\:and\: {h(x)}$. Matrix multiplication is performed on ${f(x)^T}$ and ${g(x)}$ and a SoftMax function is applied on the output. This results in generation of the attention map. The attention map is again multiplied with ${h(x)}$. A final $1\times1$ convolution operation is performed on the output, to generate ${v(x)}$. The output of the self-attention algorithm is then added to the original hidden layer through a multiplication factor of $\gamma$. Thus, the final output is given as,
\begin{equation}
    y_i = \gamma o_i + x_i
\end{equation}
$\gamma$ is a learnable variable which is initialized to 0. In the initial stages of training, the network focuses on local cues. As training progresses, more weight is given to non-local features. Thus, the network learns easier tasks first, and gradually learns complex connections with distant spatial features \cite{Zhang2019a}. The final Pix2Pix model with improvements is shown in Fig. \ref{fig:gan_architecture}

\subsection{Training Stabilization} During GAN training, we observe that the discriminator usually gets trained fairly quickly. This results in it overpowering the generator, by accurately classifying generated images as fake, thus disabling the generator from improving. The Two Time-Scale Update Rule (TTUR) presented in \cite{Heusel2018}, advocates the use of separate learning rates for the generator and the discriminator. Implementation of TTUR in SAGAN \cite{Zhang2019a} was shown to speed up the training of the discriminator. 

We propose a switching TTUR technique for efficiently training the network. Initially, we set the learning rate for the generator higher than the discriminator. This induces divergence in the GAN losses after low-level features are captured. We then switch the learning rates for the generator and the discriminator and continue training while periodically decreasing both the learning rates. We also use label-smoothing to reduce "Over Confidence" of the discriminator during loss generation \cite{Salimans2016}\cite{Pereyra2017}.

\begin{table*}[ht]\centering
  \begin{tabular}{l c c c c}
    \hline
    Experiment & SSIM (primary) & SSIM (secondary) & PSNR (primary) & PSNR (secondary) \\[0.5ex]
    \hline
    Default & 0.492 & 0.503 & 28.4780 dB & 28.9571 dB \\
    SN-Pix2Pix + SN-ESRGAN & 0.5 & 0.463 & 28.7491 dB & 28.2375 dB \\
    AASNA-Pix2Pix + SN-ESRGAN & \textbf{0.566} & \textbf{0.534} & \textbf{29.7563 dB} & \textbf{29.68949 dB} \\
    \hline
  \end{tabular}
  \caption{\label{tab:psnr}SSIM and PSNR scores for best models from different experiments}
\end{table*}

\begin{table*}[ht]\centering	
  \begin{tabular}{l c c c c c c}	
    \hline	
    Mode &  \multicolumn{2}{c}{Primary} & \multicolumn{2}{c}{Secondary}& \multicolumn{2}{c}{Tertiary}\\
    \hline
    {}          & Density & Coverage & Density & Coverage & Density & Coverage \\
    \hline	
    Default & 0.9353 & 0.7809 & 0.6966 & 0.6592 & 0.4346 & 0.6029\\	
    SN-Pix2Pix + SN-ESRGAN & 0.9162 & 0.5432 & 0.5865 & 0.4851 & 0.4084 & 0.4157\\	
    AASNA-Pix2Pix + SN-ESRGAN & \textbf{0.9527} & \textbf{0.8840} & \textbf{0.9325} & \textbf{0.8765} & \textbf{0.7822} & \textbf{0.6947}\\	
    \hline	
  \end{tabular}	
  \caption{\label{tab:d_c_comparison}Comparison b/w the three implementations on basis of Density and Coverage values for Primary, Secondary and Tertiary datasets.}
\end{table*}

\section{IMPLEMENTATION}
The following section describes the implementation details of our pipeline. We use ADAM \cite{Kingma2017} Optimizer for all the models and set the Lipschitz constant to 1 for spectral normalization.

\subsection{Data}
For image super-resolution, we use registered optical microscope and scanning electron microscope images for training the GANs. Since data extraction is a difficult and time-consuming process, a total of 578 original image pairs were available. Since these images were of 4K resolution, we extracted small 1024x1024 patches from the original images. These patches were resized to 512x512 resolution for computation efficiency.

We perform data augmentation to increase the size of the dataset, add more variability, and to prevent overfitting to the small sample size. For particle size analysis (PSA), once the images are super-resolved, we stitch the super-resolved images together to get the original sized image (4752x3168), which are then trained to detect the particles. Ground truth data for PSA is generated by human annotators. We compare our model output annotations against ground truth human annotations and other algorithms that have been proposed for particle size analysis using image processing. 

\subsection{Training}
\subsubsection{AASNA-Pix2Pix} For multi-modal image translation, we used 512x512 patches of OM and SEM pairs. Initially, the input images were read in RGB. But we found higher generalizability by the model for grayscale OM input images. We attribute this to the variability in illumination in the images captured by optical microscopes. For the Pix2Pix model with Anti-Aliasing, Spectral Normalization, and Self Attention (AASNA-Pix2Pix) learning rates of 0.005 and 0.001 were used for the generator and the discriminator respectively. We switch the learning rates at around half the epochs and then train the network with periodically decaying learning rates. For momentum, the parameters are set as $\beta_{1}=0.5, \beta_{2}=0.999$ and the RMSE loss parameter $\lambda_{RMSE}=5$.

\subsubsection{Improved-ESRGAN} For super-resolution using our ESRGAN with SN (Improved ESRGAN), input images from the AASNA-Pix2Pix model of size 512x512 were resized to 128x128. The 128x128 images were then super-resolved to get a 512x512 size image outputs using ESRGAN-SN. For training, we set learning rates to 0.003 and 0.001 for the generator and the discriminator respectively. Similar to previous training, we switch the learning rates, when the losses start to diverge and decay them periodically after the switch. We use a set of 23 Residual-in-Residual Blocks (RRDBs) \cite{Lim2017} in the generator. For momentum, the parameters are set as $\beta_1=0.9, \beta_2=0.999$.

\subsubsection{Particle Size Analysis}
The data was collected such that all practical modalities were covered. We divided the entire set of images into primary, secondary, and tertiary datasets. Ground truths (GT) were created by manually annotating the particles. 80\% of the primary dataset was used for training and the remaining for validation and testing. The distributions of all datasets are shown below.

The training data and corresponding ground-truth, before being fed to the SR-PSA network, were converted to grayscale and scaled-down by 4 and 8 to overcome scale imbalance. Feeding the network at different scales helps in extracting features of larger as well as smaller sized particles, which in turn helps in detecting particles from various size distributions. The ground-truth radii were normalized by a factor called max-radius, which is considered as a hyperparameter. The output at different scales was finally merged to get the final predictions. The objective was to optimize the Loss function given in Eq. \ref{SRPSA_loss}. Throughout the training, we used a batch size of 4 with ADAM optimizer with learning rate, weight decay, $\lambda$, and max-radius as 0.0001, 0.000001, 5, 5 respectively.

\begin{table*}[t!]\centering		
  \begin{tabular}{l c c c c c}		
    \hline		
    Experiment & Mean & D50 &  Mode & StDev & CV \\		
    \hline		
    Ground Truths & 9.34 & 8.89 & 8.60 & 2.33 & 24.96\\		
    SR-PSA         & 9.19 & 8.86 & 8.55 & 2.42 & 26.32\\		
    YOLOv5          & 9.33 & 8.89 & 8.64 & 2.35 & 25.24\\		
    \hline		
  \end{tabular}		
  \caption{\label{tab:table-primary}Comparison b/w Ground Truth, SR-PSA and YOLOv5 Predictions on Primary data set}	
\end{table*}	
\begin{table*}[t!]\centering		
  \begin{tabular}{l c c c c c}		
    \hline		
    Experiment & Mean & D50 &  Mode & StDev & CV\\		
    \hline		
    Ground Truths & 6.74 & 6.47 & 6.46 & 1.48 & 22.00\\		
    SR-PSA         & 6.94 & 6.71 & 6.47 & 1.48 & 21.38\\		
    YOLOv5          & 7.06 & 7.11 & 6.77 & 1.46 & 20.70\\		
    \hline		
  \end{tabular}		
  \caption{\label{tab:table-secondary}Comparison b/w Ground Truth, SR-PSA and YOLOv5 Predictions on Secondary data set}	
\end{table*}	
\begin{table*}[t!]\centering		
  \begin{tabular}{l c c c c c}		
    \hline		
    Experiment & Mean & D50 &  Mode & StDev & CV\\		
    \hline		
    Ground Truths & 6.73 & 6.86 & 7.01 & 0.92 & 13.62\\		
    SR-PSA         & 6.83 & 6.88 & 6.97 & 0.65 & 9.49\\		
    YOLOv5          & 7.6 & 7.11 & 7.35 & 1.3 & 17.08\\		
    \hline		
  \end{tabular}		
  \caption{\label{tab:table-tertiary}Comparison b/w Ground Truth, SR-PSA and YOLOv5 Predictions on Tertiary data set}	
\end{table*}	

\begin{table*}[t!]\centering		
  \begin{tabular}{l c c c}		
    \hline		
    Particle Count & GT & SR-PSA (\% detected) &  YOLOv5 (\% detected)\\
    \hline		
    Primary           & 47166 & 46282 (98.13 \%) & 45765 (97.03 \%)\\
    Secondary         & 14806 & 14772 (99.77 \%) & 11689 (78.95 \%)\\
    Tertiary          & 14697 & 13397 (91.15 \%) & 8040 (54.71 \%)\\
    \hline		
    Combined Detections    &  -    & 97.10 (\%) & 85.42 (\%)\\		
    \hline		
  \end{tabular}		
  \caption{\label{tab:table-count}Comparison of Particle detection's b/w SR-PSA and YOLOv5 Predictions with respect to GT (ground truths) on combined data set (primary+secondary+tertiary)}	
\end{table*}

\section{EXPERIMENTS} 
We use separate evaluation metrics for qualitative measurement of GAN outputs and quantitative measurement of the PSA outputs. 

\subsection{Super Resolution} Evaluation of the quality of synthetic images is a highly domain-specific question. One of the most intuitive ways for qualitative measurement of generated images is manual GAN evaluation. Manual evaluation is suitable for the development and quick prototyping of models by changing hyperparameters. However, due to inherent bias in human supervision and large amount of time involved, it is not suitable. Hence, we rely on qualitative measures to evaluate the performance of our GANs. For all comparison metrics, we use three sets of implementations. For each implementation, the best model obtained after hyperparameter tuning was used to get the evaluation metrics.

\subsubsection{Default Implementation} Initially default implementations of Pix2Pix and ESRGAN were replicated individually for our use-case. While ESRGAN is powerful, it fails to perform end-to-end multi-modal super-resolution effectively. Pix2Pix model is an image translation cGAN which uses U-NET architecture for multi-modal transfer of peculiar image features. A standalone Pix2Pix model trained on OM-SEM pairs performs image translation of the particles and does not solve the problem of thick and unsharp edges in OM images. Hence, we combined Pix2Pix with ESRGAN to obtain truly super-resolved particles from the input OM images.

\subsubsection{GANs with Spectral Normalization} We observed frequent non-convergence of the models in the default implementations. We attribute the non-convergence problem to the discriminator learning faster than the generator, resulting in a sub-optimal generator residing in local minima, which keeps generating identical particles without any diversity. To bring stability to the training of both the models, SN was used in their discriminators \cite{Miyato2018}. With the addition of SN and RMSE loss for the Pix2Pix model, the circularity of particles increased, however, these changes resulted in the images looking less realistic. 

\subsubsection{AASNA-Pix2Pix with ESRGAN-SN} Inspired by the Self-Attention GAN (SA-GAN) paper \cite{Zhang2019a}, we added self-attention to our generator to capture spatially relevant information. SA-GAN implements spectral normalization in generator and discriminator and shows significant improvement in image generation capabilities of GANs. For a cGAN use-case, we implement self-attention in the up-sampling part of the network so that the generator doesn't focus only on the local features. This enables the generated images to have consistent texture information even in the spatially distant parts. 

\begin{figure}[h]	
    \centering	
      \begin{subfigure}[b]{0.3\linewidth}	
        \includegraphics[width=\linewidth]{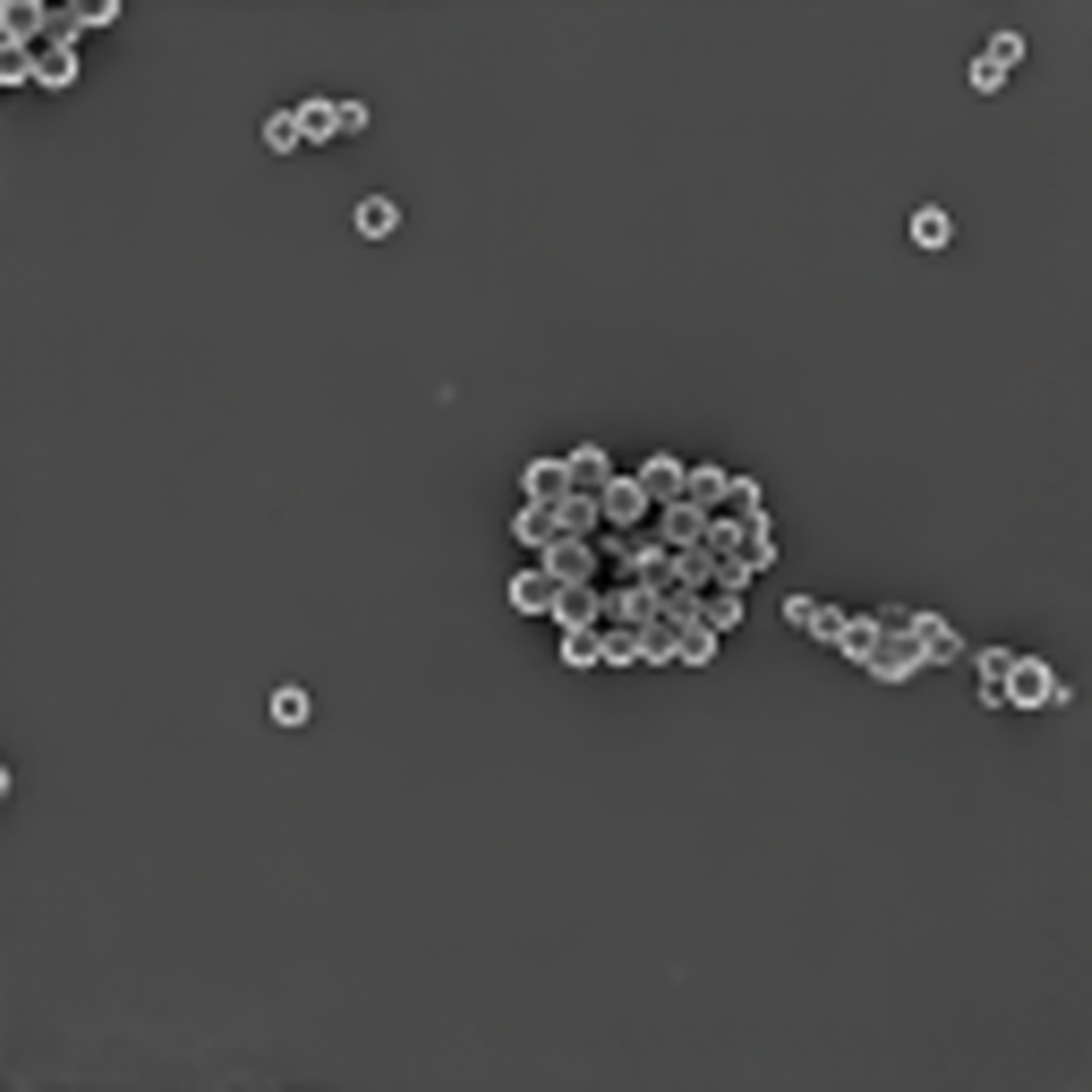}	
        \caption{L2 Loss}	
        \label{fig:l2loss}	
      \end{subfigure}	
      \begin{subfigure}[b]{0.3\linewidth}	
        \includegraphics[width=\linewidth]{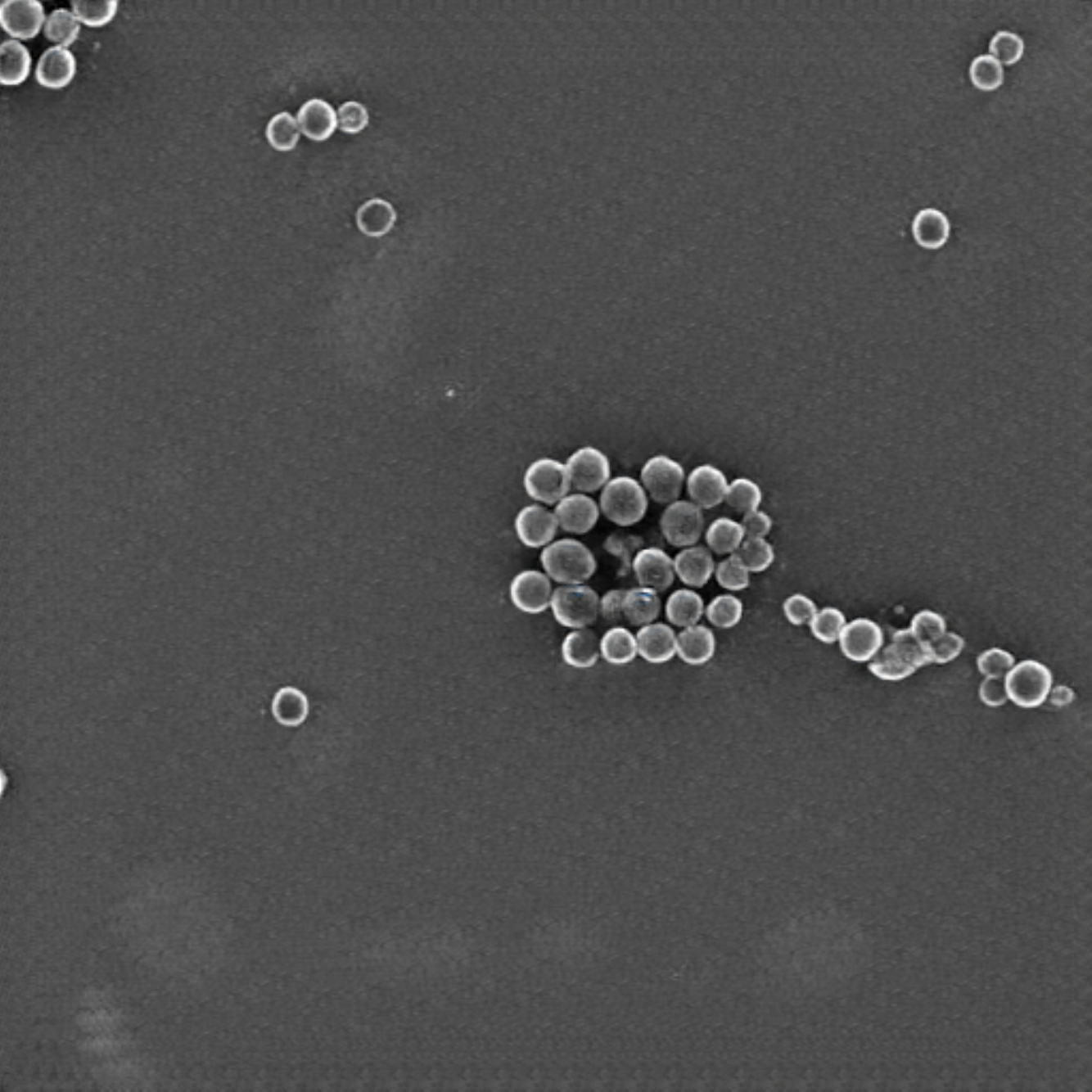}
        \caption{L1 Loss}	
        \label{fig:l1loss}	
      \end{subfigure}	
      \begin{subfigure}[b]{0.3\linewidth}	
        \includegraphics[width=\linewidth]{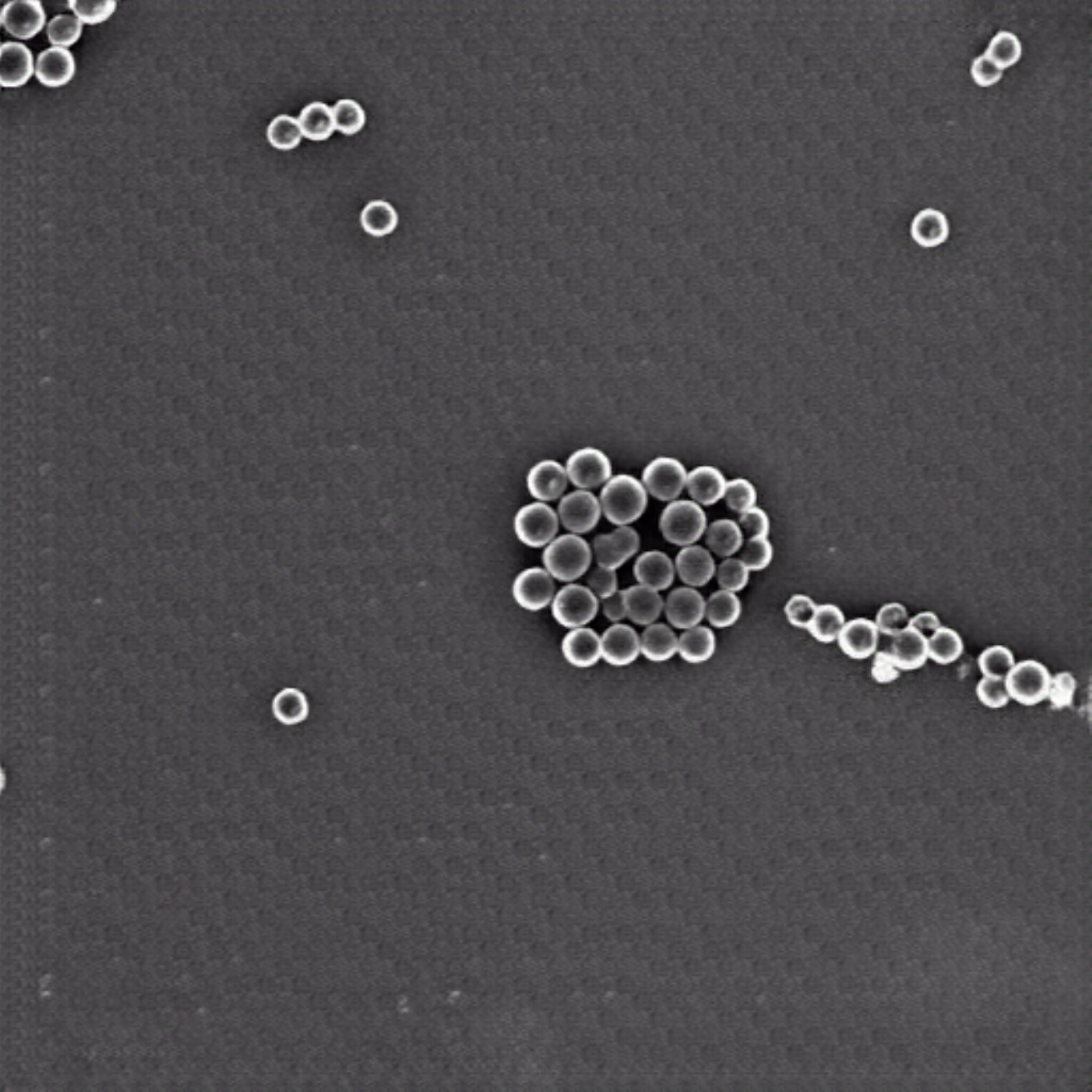}
        \caption{RMSE Loss}	
        \label{fig:rmseloss}	
      \end{subfigure}
    \caption{(a) L2 loss result shows significant blurring at the boundaries of the particles. (b) L1 loss result shows sharp, but non-circular boundaries. (c) RMSE loss result shows slightly thicker, but highly circular boundaries.}	
    \label{fig:l1vsl2vsrmse}	
\end{figure}

To further bring stability to our models, we use anti-aliasing pooling instead of max-pooling. We use a normalized Binomial-5 filter which has shown to increase the consistency in CNNs \cite{Zhang2019}. Since ESRGAN only looks at the Pix2Pix output, we add self-attention and anti-aliasing to the Pix2Pix model. This enables Pix2Pix to generate high quality images consistently irrespective of the aberrations in the input, thus maintaining the quality of the final results. Hence, the bulk of the responsibility is handled by AASNA-Pix2Pix, so that the task becomes easier for the improved ESRGAN to accomplish. Results of experiments with different generator loss functions are shown in Fig. \ref{fig:l1vsl2vsrmse}.

\subsection{Evaluation}
In the community, standard GAN performance is usually benchmarked using metrics like Inception Score \cite{Zhang2019}, Fréchet Inception Distance (FID) \cite{Heusel2018}, Structural Similarity Index Measure (SSIM) \cite{Wang2004} and Peak Signal-to-Noise Ratio (PSNR). For standard GANs, since the real data is not exactly identical to the generated data, one-to-one metrics like SSIM and PSNR do not work well. However, for conditional GANs, if real images corresponding to the input are available, these metrics are often used to measure the quality of generated images.

\subsubsection{Structural Quality Metrics} PSNR was originally used as a quality measurement tool between compressed and reconstructed images. We use PSNR to show marginal improvements in the image quality when compared to the ground truth, as our implementation got better. We also compare our implementations using SSIM. An empirical comparison between the two methods is given in \cite{Hore2010}. We show that our AASNA-Pix2Pix + improved ESRGAN results are better than the default implementation as shown in Table \ref{tab:psnr}.

\subsubsection{Fidelity and Diversity Metrics} Besides structural similarity, a good metric for GAN outputs is FID. However, FID scores do not make sense for our application because the generated images are quite similar looking to the real images. Hence, singular FID scores of less than 2 are obtained, which does not prove a reliable metric to compare. Recently, precision and recall based metrics have been proposed by \cite{Sajjadi2018},\cite{Kynkaeaenniemi2019}. These metrics are based on an implicit quality of the data called "manifolds". For complex high-dimensional data, that has some underlying geometric differences, the differences can be captured by its manifold in a 2D space. To evaluate our model, we use the recently developed metrics of density and coverage \cite{Naeem2020}, which are an improvement over the aforementioned precision and recall.

\begin{figure}[h]	
    \centering	
    
      \begin{subfigure}[b]{0.9\linewidth}	
        \includegraphics[width=\linewidth]{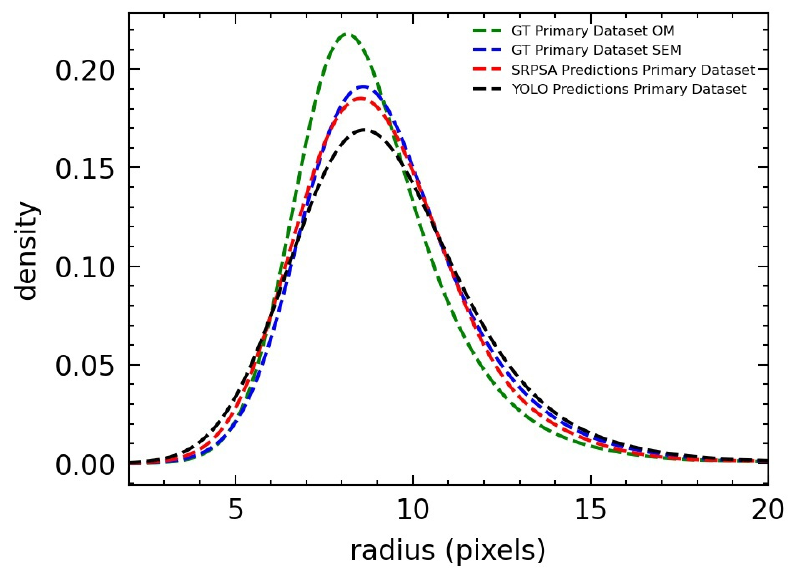}
        \caption{Results for Primary Dataset}
        \label{fig:comp_prim}
      \end{subfigure}	
      
      \begin{subfigure}[b]{0.9\linewidth}	
        \includegraphics[width=\linewidth]{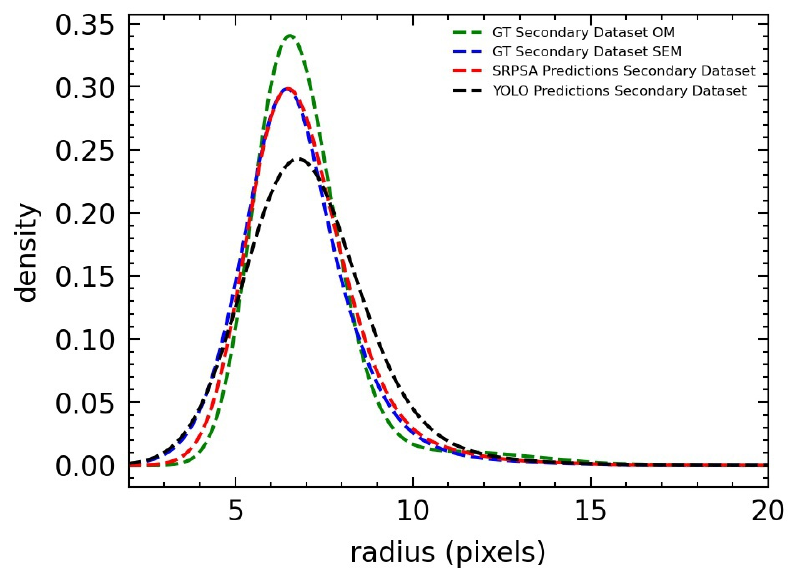}
        \caption{Results for Secondary Dataset}
        \label{fig:comp_sec}
      \end{subfigure}
      
      \begin{subfigure}[b]{0.9\linewidth}	
        \includegraphics[width=\linewidth]{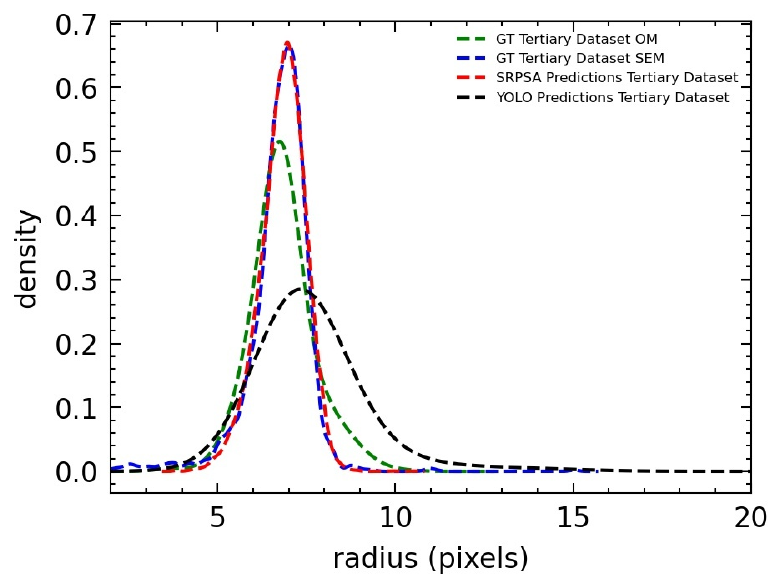}
        \caption{Results for Tertiary Dataset}
        \label{fig:comp_tert}
      \end{subfigure}
      
    \caption{Comparison between ground truths for OM and SEM datasets, and predictions by the SR-PSA and YOLOv5 models for (a) Primary Dataset, (b) Secondary Dataset, and (c) Tertiary Dataset.}		
\end{figure}

For calculation of manifolds, representative embeddings are needed for the ground-truth and the generated images. Many evaluation metrics rely on the use of models pre-trained on ImageNet. This may affect the evaluation as the pre-trained models include a dataset bias as reported in \cite{Geirhos2019},\cite{Naeem2020}. Thus, we use a randomly initialized Inception\_V3 model to obtain the representative embeddings, as random embeddings are free from any kind of bias and hence can provide sensible evaluation metrics as reported in \cite{Naeem2020}. 

To test the density and evaluation metrics for our use case, we generated two sets of synthetic images. One set of images had random backgrounds with particles intact, and the other set had washed out particles with the background intact. For the 1st set, with particles intact, we observed a high coverage value. For the 2nd set, with the background intact, we observed a high density value. Thus, for comparison between different models, we use the density and coverage values as indicators of the quality of the background and the generated particles respectively. Values for the three datasets are presented in Table \ref{tab:d_c_comparison}.

We observe a decline in the results of the second implementation with just spectral normalization, as the images generated had many artifacts in the background for secondary and tertiary testing datasets. With the implementation of Anti-Aliasing and Self-Attention, we see a significant improvement across all the datasets in both, density as well as coverage.

\subsection{Particle Size Analysis} The important parameters in PSA are mean, d50, mode, coefficient-of-variation (CV) and standard deviation (StDev) of the particle size distribution. YOLOv5 object detection model was trained on the same primary dataset to have a fair comparison with SR-PSA. Since the testing dataset distributions were completely different from the training, our evaluation metrics determine the generalizability of SR-PSA. To quantify the robustness of our model, we compare the kernel density estimation (KDE) plots of SR-PSA and the YOLOv5 model with the human-generated annotations ("Ground Truth") on OM and SEM images. This comparison was done for the primary, secondary and tertiary datasets. 

The distributions were made in a non-parametric way using the Improved Sheather-Jones (ISJ) algorithm. ISJ estimates bandwidth by minimizing the asymptotic mean integrated square error (AMISE) \cite{Botev2010}.	
From the distribution the statistics such as mean, median (D50), mode, StDev, CV were estimated. The comparison of distributions for primary, secondary and tertiary data sets are shown in Fig. \ref{fig:comp_prim}, Fig. \ref{fig:comp_sec}, and Fig. \ref{fig:comp_tert} respectively. As can be seen from the plots, SR-PSA predictions have consistently followed the same distribution as the manual annotations on the SEM images.
	
\begin{figure}[h]	
    \centering	
      \begin{subfigure}[b]{0.45\linewidth}	
        \includegraphics[width=\linewidth]{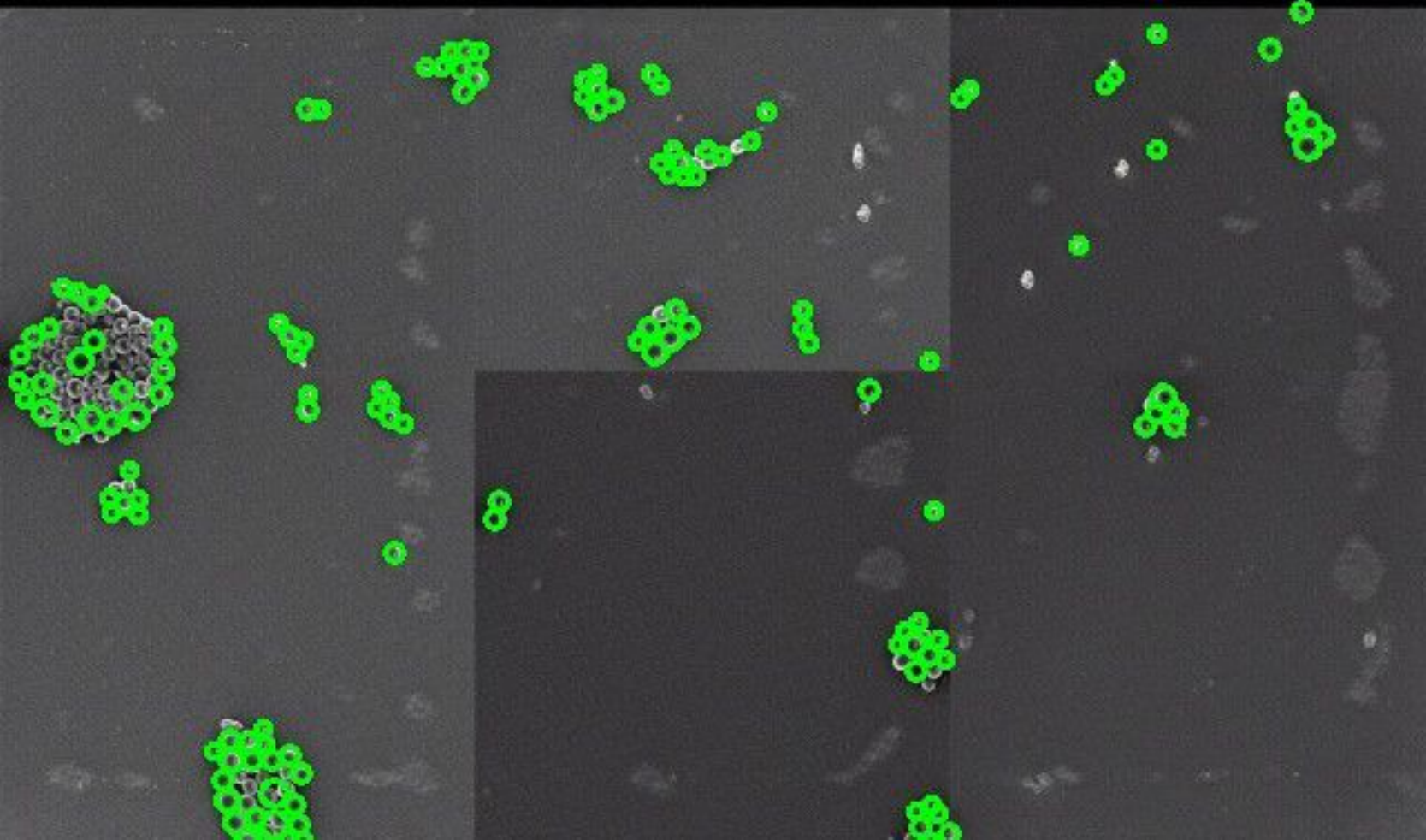}	
        \caption{YOLOv5}	
        \label{fig:psayolo_1}	
      \end{subfigure}	
      \begin{subfigure}[b]{0.45\linewidth}	
        \includegraphics[width=\linewidth]{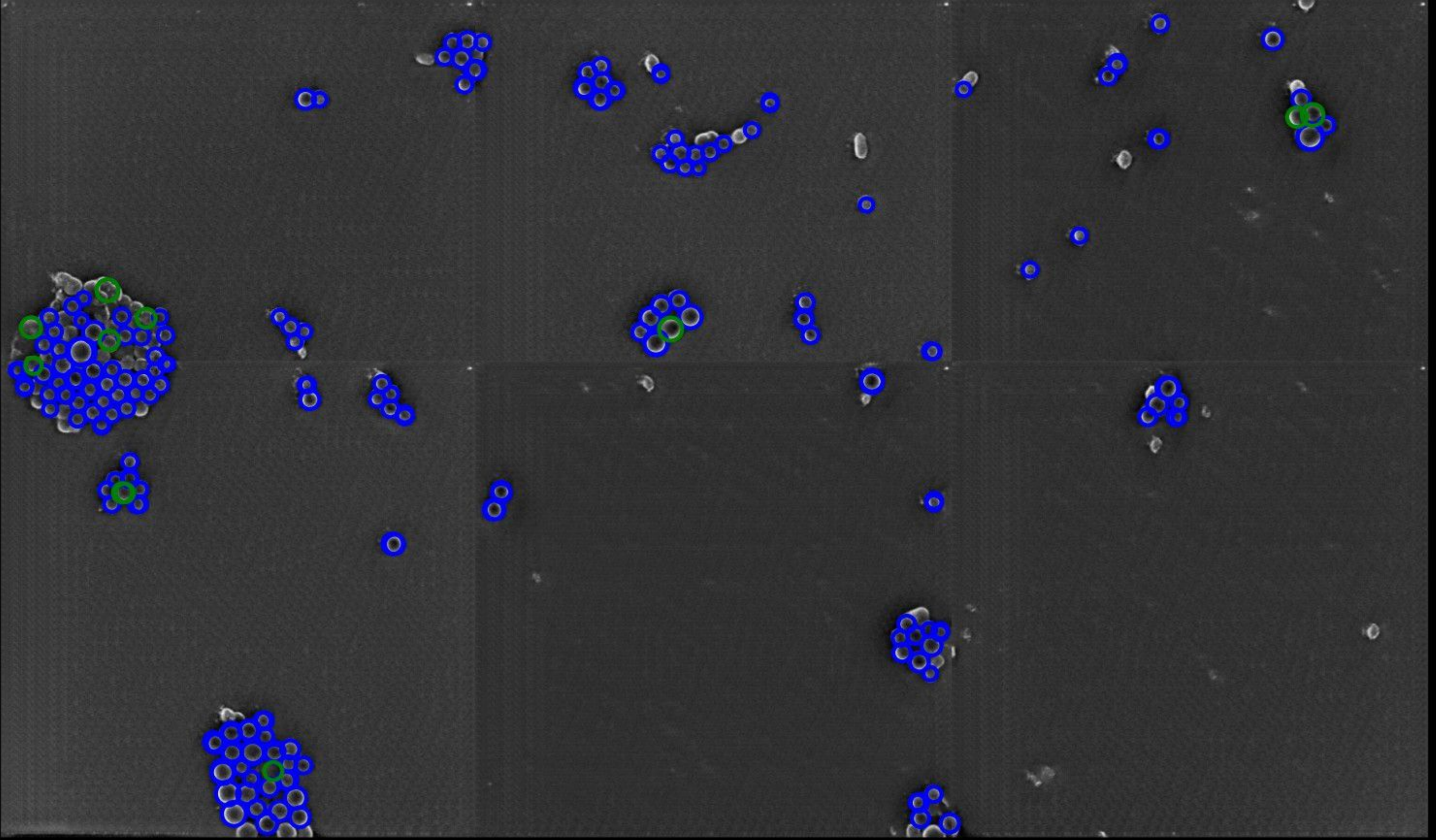}	
        \caption{Our Network}	
        \label{fig:srpsa_1}	
      \end{subfigure}	
      \begin{subfigure}[b]{0.45\linewidth}	
        \includegraphics[width=\linewidth]{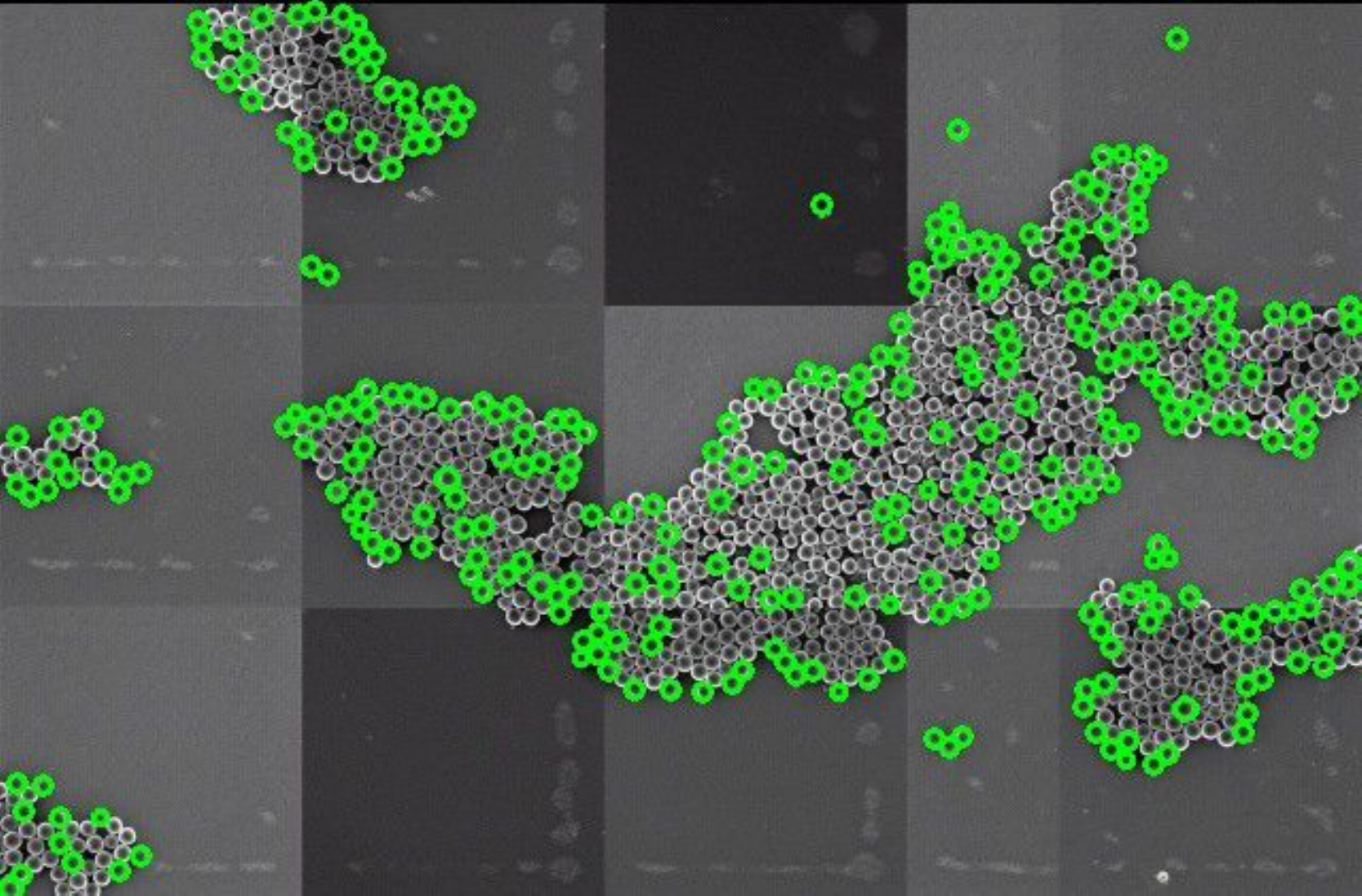}	
        \caption{YOLOv5}	
        \label{fig:psayolo_2}	
      \end{subfigure}
      \begin{subfigure}[b]{0.45\linewidth}	
        \includegraphics[width=\linewidth]{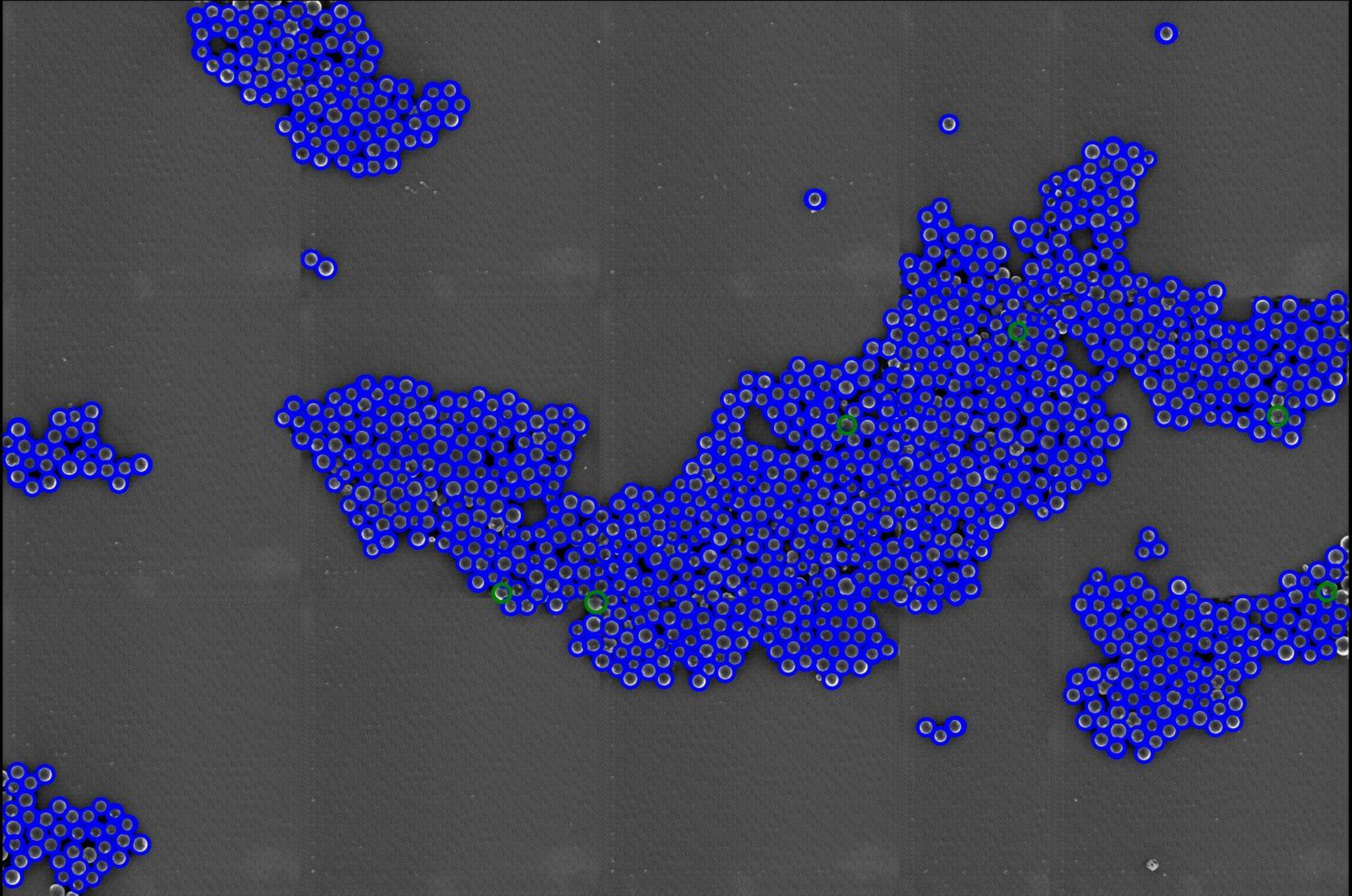}	
        \caption{Our Network}	
        \label{fig:srpsa_2}	
      \end{subfigure}
    \caption{(a) \& (b) Predictions on Secondary Dataset, (c) \& (d) Predictions on Tertiary Dataset}	
    \label{fig:sec_yolo_vs_srpsa}	
\end{figure}
	
The comparison of statistics for primary, secondary and tertiary data sets are shown in Table \ref{tab:table-primary}, Table \ref{tab:table-secondary} and Table \ref{tab:table-tertiary} respectively. Table \ref{tab:table-count} shows detection capabilities of SR-PSA and YOLOv5 opposed to the ground truth annotations. Along with the statistical accuracy, the amount of particles detected using SR-PSA is significantly higher than the YOLOv5 model. YOLOv5 performed well for the primary dataset but failed to work on secondary and tertiary data sets with a different particle size distribution. Fig. \ref{fig:sec_yolo_vs_srpsa} shows the inability of YOLOv5 to detect particles that are smaller and densely packed. On the other hand SR-PSA, consistently detects particles on distributions that are drastically different from the primary dataset.

\section{CONCLUSION AND FUTURE WORK}
In this paper, we propose a powerful reproducible method for super-resolution of optical microscopic particles and consequent particle size analysis. In addition, we improve on the state-of-the-art image-based cGAN methods to present the AASNA-Pix2Pix architecture for image translation. The combination of GANs in our method, which delivers high fidelity multi-modal super-resolution can be extended to other domains with similar requirements too. We demonstrate the application of our particle sizing algorithm on top of the super-resolved images that achieves accurate near-human performance for particle size analysis on circular particles. 

Further investigation is still needed to understand the generalizability of the proposed pipeline across various industries. Moreover, we restricted the use of self-attention to the three smallest layers in our Pix2Pix model. More experiments are needed with other spatial attention techniques to reduce the computational complexity of the model, whilst maintaining the quality of the performance.

{
\renewcommand{\clearpage}{}
\bibliographystyle{ieeetr} 

}
\begin{IEEEbiography}[{\includegraphics[width=1in,height=1.25in,clip,keepaspectratio]{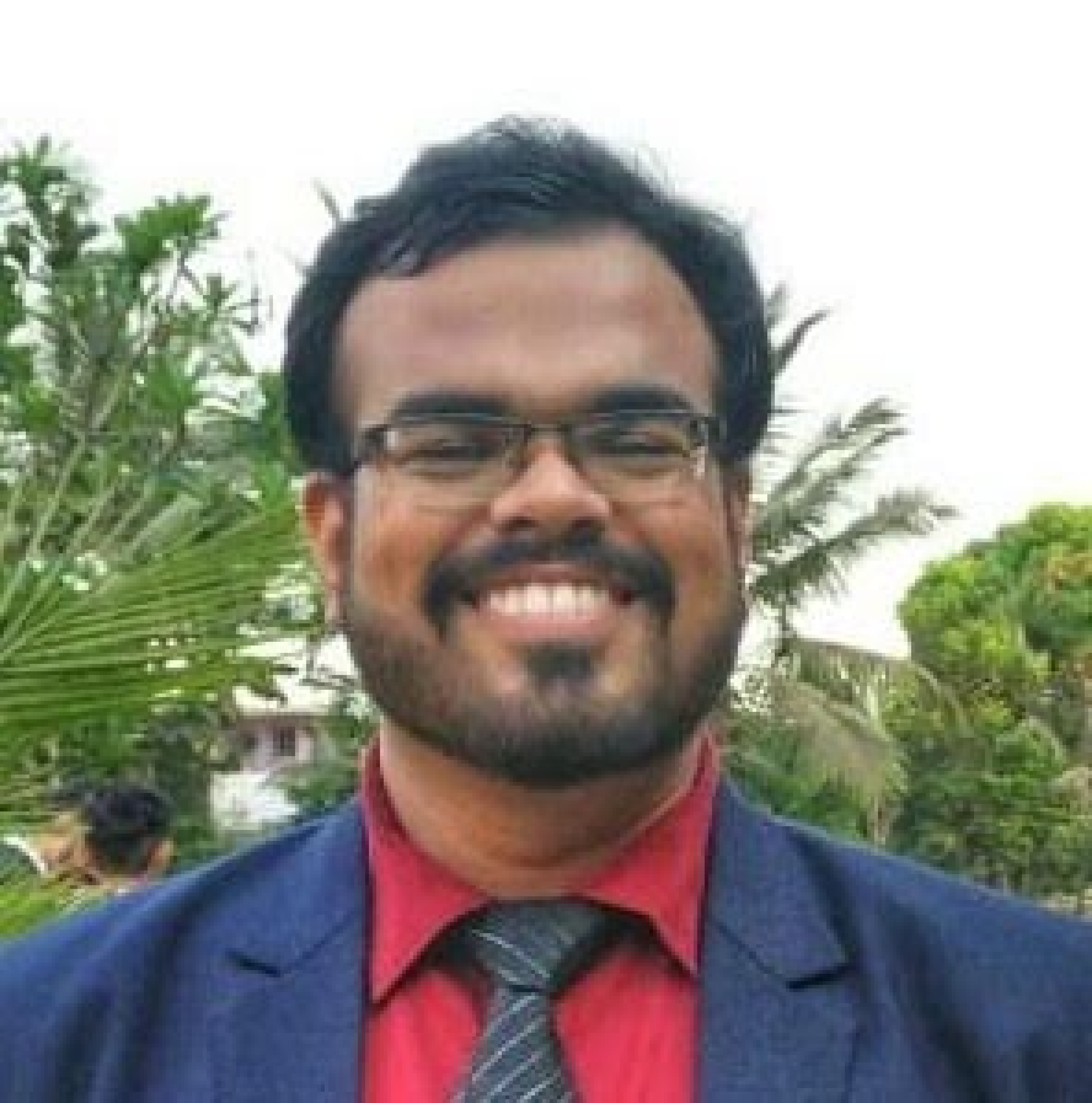}}]{Sarvesh Patil}
is a data scientist at HyperWorks Imaging. He received his Bachelor's Degree in Electronics and Telecommunication from the VES Institute of Technology, Mumbai. His research interests lie in computer vision, with a focus on object detection, generative adversarial networks, and scene understanding.
\end{IEEEbiography}

\begin{IEEEbiography}[{\includegraphics[width=1in,height=1.25in,clip,keepaspectratio]{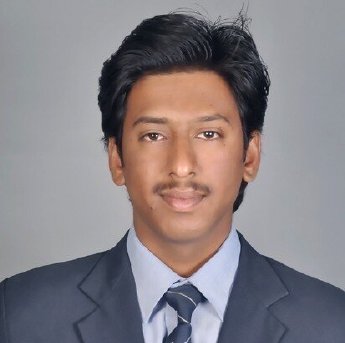}}]{Chava Y P D Phani Rajanish}
received his B.Tech in Aerospace Engineering from UPES,  M.Tech in Thermal Engineering from KL University, and is pursuing his Masters in Data Science from BITS Pilani. He is currently working at Hyperworks Imaging Pvt Ltd, Bangalore, India as a Data Scientist with a focus on Computer Vision and Data Engineering.

\end{IEEEbiography}

\begin{IEEEbiography}[{\includegraphics[width=1in,height=1.25in,clip,keepaspectratio]{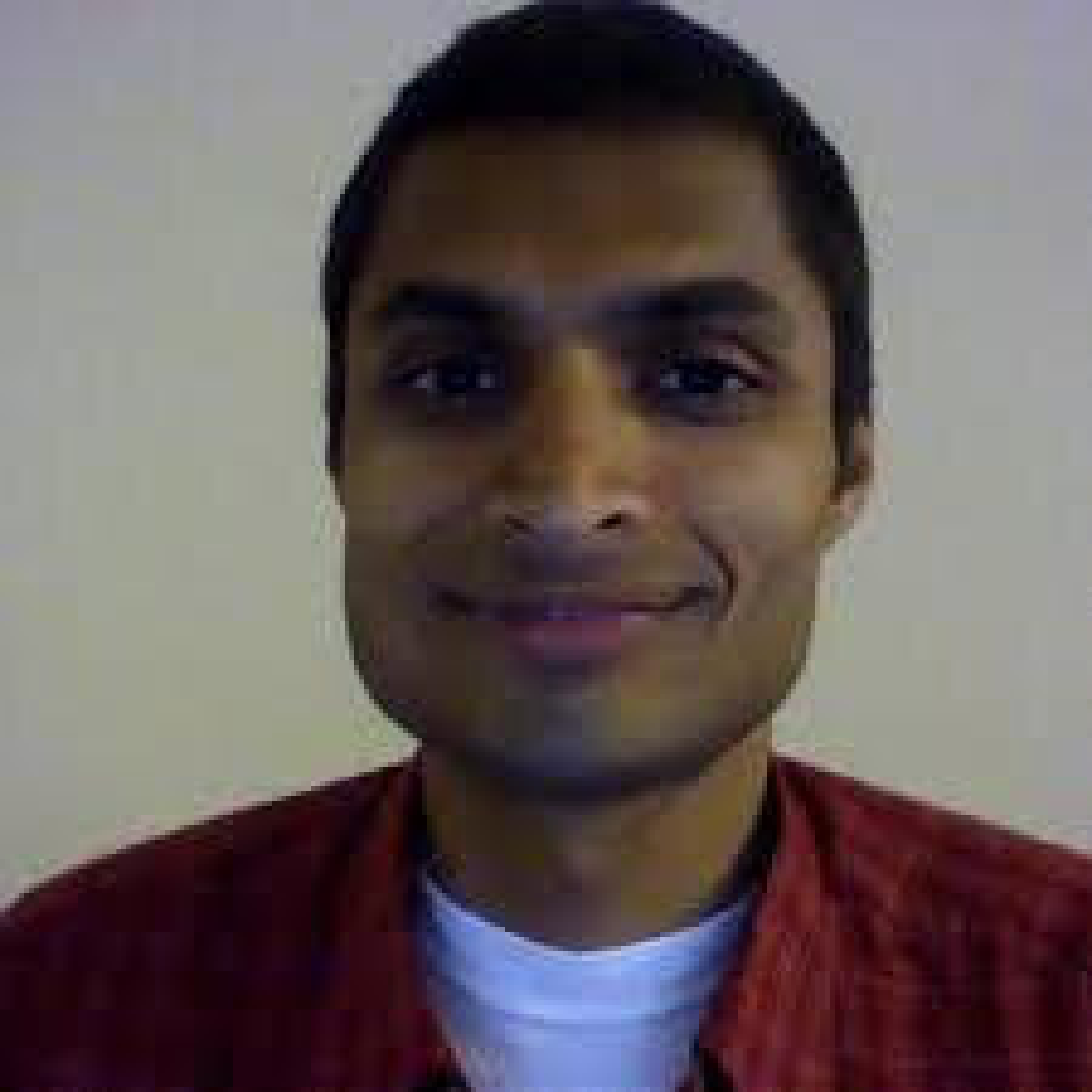}}]{Naveen Margankunte}
received his PhD in physics from the University of Florida (UF). He is currently CEO of HyperWorks Imaging enabling creation and deployment of bespoke AI and ML solutions for various domains including advanced R\&D, Agtech, and Retail.

\end{IEEEbiography}
\vfill

\end{document}